\documentclass{JHEP3} 
\usepackage{amsmath,amssymb,amsfonts,nicefrac}
\usepackage{graphicx}

\newcommand{\oao}[2]{{#1\atopwithdelims[]#2}}

\def\zi{\mathbb{Z}}

\def\di{\text{d}}

\def\slr{\text{SL(2},\mathbb{R})}
\def\slc{\text{SL(2},\mathbb{R})/\text{U}(1)}

\def\rrangle{\rangle \hskip-.5mm \rangle}

\def\p{\partial}
\def\pb{\bar{\partial}}
\def\cn{\mathcal{N}}

\title{Rolling Tachyon in Anti-de Sitter Space-Time}

\author{Dan Isra\"el$^{1}$  and 
Eliezer Rabinovici$^{1,2}$
\\
{\small \it $\, ^1$ Racah Institute of Physics, The Hebrew University \\
Jerusalem 91904, Israel \\
$\, ^2$ Department of Physics, 
Theory Division,\\
CERN, CH-1211 Geneva 23,
Switzerland }
}

\abstract{
We study the decay of the unstable D-particle in 
three-dimensional anti-de Sitter space-time using worldsheet 
boundary conformal field theory methods. We test the open string completeness 
conjecture in a background for which the phase space available is only 
field-theoretic. This could present a serious challenge to the claim. We compute the emission 
of closed strings in the AdS$_3 \times$ S$^3 \times$ T$^4$ 
background from the knowledge of the exact corresponding 
boundary state we construct.  
We show that the energy stored in the brane is mainly converted 
into very excited long strings. The energy stored in short 
strings and in open string pair production is much smaller and finite for 
any value of the string coupling. We find no "missing energy" problem. We compare 
our results to those obtained for a decay in flat space-time and to a background in 
the presence of a linear dilaton. Some remarks on holographic aspects of the problem are 
made. }

\preprint{RI-00-08-06\\
CERN-PH-TH/2006-161\\
hep-th/0609087}

\begin{document}

\section{Introduction and summary} 
Understanding the physics of time-dependent backgrounds is one of the the greatest challenges in 
string theory. It is important for the understanding of basic issues in quantum gravity and 
for obtaining some hints of stringy effects in primordial cosmology.  The current 
technology, based on a topological expansion 
of string worldsheets in a first quantized framework, leaves  
many problems unsolved, such as  
understanding the analytical continuation to Lorentzian signature 
and backreaction that arises in time-dependent space-times. 
Some exact two-dimensional conformal field theories with 
a time-dependent target space-time were analyzed~\cite{Kounnas:1992wc,Nappi:1992kv,Cornalba:2002fi,Nekrasov:2002kf,
Elitzur:2002rt}. It is not easy to extract physical information from them, in particular 
due to the presence of singularities in the correlators 
in some cases~\cite{Liu:2002kb,Berkooz:2002je,Giveon:2004rw}, that  
may signal a large back-reaction on the
geometry~\cite{Horowitz:2002mw,Pioline:2003bs} 
and the breakdown of perturbation theory (see however~\cite{Cornalba:2003kd}).

Time-dependent backgrounds also play an important role in 
the search for new string vacua in the string background landscape. 
Detecting and following infrared instabilities in space-time and on the worldsheet 
has been a tool for such  investigations. 
The more risky expeditions were those following the trajectories of closed
string tachyons (see for example~\cite{Elitzur:1991cb}).  Following open
string tachyons (see~\cite{Sen:1998sm,Elitzur:1998va}, the review~\cite{Sen:2004nf} and 
references therein) and localized closed string tachyons (see for example~\cite{Adams:2001sv,Barbon:2001di}) 
is somewhat more tractable and also of great interest.

In this paper we study some aspects of a particular time-dependent process involving the
decay of unstable D-branes, described by an integrable boundary deformation of the string 
worldsheet~\cite{Sen:2002nu,Gutperle:2003xf}. 
The main simplification is that the minimum of the tachyon potential, the closed string 
vacuum, is known~\cite{Sen:1999nx}. The study of open string tachyon condensation is not 
only important to unveil fundamental aspects 
of string theory dynamics but also to attempt to build 
realistic cosmological scenarios in
string theory~\cite{Kachru:2003sx,Cremades:2005ir}.

The tree-level computation of closed string production leads to an exponentially divergent emission 
of highly massive non-relativistic excited
strings~\cite{Lambert:2003zr,Gaiotto:2003rm}. 
This divergence is unphysical for any 
non-zero value of $g_s$ (because the initial energy available is finite 
and equal to the D-brane mass), and indicates in that case inherent difficulties in a
perturbative analysis. Considerations following from the space-time effective action  
also suggest that the remnant of the process is a pressure-less 
``tachyon dust'', whose nature was studied in~\cite{Sen:2002nu,Sen:2002in} and
characterized there as somewhat mysterious. The main issue addressed was if the
open string description in terms of tachyon dust can be reformulated in terms
of a closed string description. If it would turn out that the closed string
component of the decay can't account for all the energy available in the 
D-brane before its decay than a "missing energy" puzzle would emerge. It would not be
clear what type of matter constitutes the reminder of the decay products of the
D-brane. The formulation of the question is actually more involved. In the tree-level 
approximation the mass of the decaying D-brane is infinite. If the tree-level 
approximation for the energy deposited in closed strings is finite and does not
depend on $g_s$ (\textsc{uv} finite) then a puzzle emerges, because 
the first estimate of the finite energy density of the brane would be that it is of order $1/g_s$.

In this paper we examine the "missing energy" issue in the circumstances
where that phase space is minimized, that is the case of  AdS$_3$ where the actual
entropy is not Hagedorn like  but only field-theoretic. We mainly 
consider the decay of an unstable D0-brane in this background. As the amount of
closed string radiation is also proportional to the phase space available for the closed
strings,  one  may expect this case to be a most severe challenge to the
{\it open string completeness} conjecture of Sen \cite{Sen:2003iv}, which states that the 
open string field theory description of the tachyon decay (approximated by the tachyon 
effective action) captures all the phy\-sics of the decay process, in particular 
that  it gives a ``holographically dual'' description of closed string radiation, 
even when closed string perturbation theory breaks down. We find that
in fact the outcome of the computation will be qualitatively similar to brane decay in flat
space-time. In the tree approximation the energy stored
in the closed strings is much too high and one must identify  a mechanism to
decrease the amount of energy carried by the closed strings. The open string
picture on the other hand remains rather simple.

String theory in AdS$_3$ with \textsc{ns-ns} fluxes is described by the (universal cover of the) 
$\slr$ \textsc{wzw} model~\cite{Balog:1988jb}; 
it is a solvable conformal field theory. The study of the associated boundary conformal field theory 
allowed to construct several types of $\slr$-preserving and symmetry-breaking
D-branes, first in the related Euclidean H$_3^+$ 
model~\cite{Ponsot:2001gt,Lee:2001gh} and then in the case of AdS$_3$ with Lorentzian
signature~\cite{Israel:2005ek}. Among the symmetry-breaking D-branes 
is a D-particle sitting at the origin of global coordinates~\cite{Quella:2002ns}. 
In this work we consider the non-\textsc{bps} D0-brane in
type \textsc{iib} superstrings on AdS$_3 \times$ S$^3 \times$ T$^4$ built 
out of it. We study its spectrum of fluctuations (open string modes)
that contains a tachyon. 

The presence of this open string tachyon, and the particular structure of the 
boundary state associated with this D-particle, allows one to find 
the exact boundary state describing the time-dependent 
development of this instability. The boundary deformation that 
is turned on  has the form (in the bosonic case) 
$\oint_{\p \Sigma}\di \ell \ \exp (\nicefrac{X^0 (\ell)}{\sqrt{\alpha'}})$, 
i.e. the same as one would obtain for an unstable D-brane  
in flat space-time, although the boundary state is different.

There are two (related) important differences though between the 
flat space-time case and the current analysis in AdS. First, 
the scaling of space-time energy with the oscillator number of the string   
is quite different (i.e. $E_\text{AdS} \sim N$ vs. 
$E_\textsc{flat} \sim \sqrt{N}$), meaning that the number of string 
states for a given energy is smaller. As a consequence, one may envisage that 
the production of very excited closed strings could be 
highly suppressed relative to flat space-time.

A second striking difference with the rolling tachyon in flat space-time 
is the existence of ``long strings'', i.e. macroscopic circular strings 
whose radius grows linearly w.r.t. AdS global time, while having 
a finite  energy because their (infinite) mass is balanced by the coupling to the 
\textsc{ns-ns} two-form~\cite{Seiberg:1999xz,Maldacena:2000hw}. We find that they are 
copiously produced by the brane decay, in such a way that an unrestricted sum over 
the winding sectors would give a divergent energy for the closed string radiation.  
This is obtained by computing the annulus amplitude for the rolling tachyon boundary state, 
whose imaginary part gives the mean number of emitted closed strings~\cite{Karczmarek:2003xm}.
 
Non-perturbative physics does provide a cutoff on the magnitude of the winding number $w$ 
(which has to be smaller than the number of background fundamental strings) which 
manages to regularize both the mean number of emitted closed strings and all moments of the 
emitted energy. However it is not stringent enough to give a correct estimate of the radiated energy. 
The finite energy obtained is of order  $\nicefrac{1}{g_s^2}$ while the total energy stored in 
the brane is only of order $\nicefrac{1}{g_s}$. This finite result consists of important 
contributions of long string states whose energy is larger than $\nicefrac{1}{g_s}$. Such 
contributions, although finite, are outside the realm of a reliable perturbative computations. 
As the energy stored in closed strings is larger than the available energy one should try 
to understand how the actual amount of energy is of order $\nicefrac{1}{g_s}$.
If for some reason only closed strings with energies smaller 
than $\nicefrac{1}{g_s}$, i.e. strings whose energy we can reliably calculate, 
would contribute significantly to the computation of the mean radiated energy,  
a fully satisfactory result would be obtained.

Another way to regularize the high-energy divergence of closed strings 
radiation that was advocated in the past is to consider non-critical string 
theories~\cite{Karczmarek:2003xm}. Such string backgrounds have a higher Hagedorn 
temperature compared to critical strings 
in flat space-time, therefore their ultraviolet behavior is somewhat softer. 
Previous experience suggests that anti-de Sitter space-time is the best 
possible regulator because its density of states is the same as 
that of a field theory. Therefore we do not expect closed string 
emission by a D-brane decay in linear dilaton backgrounds to be 
more suppressed at high energies than in AdS. Analyzing some examples of 
rolling tachyon in non-critical superstrings we show that closed string emission, 
computed at tree level, 
is plagued with the same divergences. Thus in both examples, AdS and linear dilaton, 
string theory manages cleverly to dissipate all the energy stored in the brane.

We compute the rate of open string pair creation~\cite{Gutperle:2003xf} by the rolling tachyon in AdS$_3$. 
We show that it is exponentially suppressed at high energies. 
Therefore perturbative string theory provides a valid description of the open string side of the 
process. These results are in agreement with Sen's hypothesis. The effective open string description 
in  flat space-time and AdS$_3$ share some similarities; the tachyon effective action 
--~which has the same domain of validity as in flat space-time if 
we choose the radius of AdS to be large w.r.t. the string length~-- predicts similar dynamics. However, 
the worldsheet \textsc{cft} description  
shows that while in the former case the mean number of emitted open strings 
is only power-like decreasing at high energies, in the latter example the convergence is exponential-like. 
The closed string side of the story seems very 
different. Instead of a ``dust'' of non-relativistic  massive closed strings, one gets a "ring" of 
long strings with very large winding number expanding at a common constant speed.

A different route to time-dependence in string theory is to use the holographic 
dualities between gravitation and gauge theory, whose 
hallmark is the AdS/\textsc{cft} correspondence~\cite{Maldacena:1997re}. 
For instance the decay of an unstable sphaleron 
in the gauge theory has been considered in~\cite{Peeters:2004rd}, 
after the proper AdS$_5$/ \textsc{cft}$_4$ identification of the 
static configuration has been found in~\cite{Drukker:2000wx}. However 
the field theory computations were compared to the flat 
space rolling tachyon solution for lack of knowledge  
of the AdS$_5 \times$ S$^5$ worldsheet theory. The example 
studied in this paper, D-particle decay in AdS$_3$, 
is a very close analogue of this setup with the advantage of  having a 
better control on the AdS side of the correspondence. 

By analogy with the AdS$_5 \times$ S$^5$  example we expect 
that the D-particle in AdS$_3$ is dual to a sort of sphaleron of the space-time
\textsc{cft} associated with the instanton dual to the D-instanton in anti de-Sitter. 
However the space-time \textsc{cft} is notoriously hard to study because 
it is singular in the regime where the bulk theory is solvable, i.e. without Ramond-Ramond fluxes. 
We leave as an open problem the precise description of the D0-brane dual in the 
space-time theory. This would help to understand the physics of 
the decay at a non-perturbative level, in particular to understand 
better the regularization of long strings production.

The production of long strings is one of the features that render the 
holographic interpretation complicated, because it decreases the central charge of the space-time 
conformal field theory. The latter is the product of two \textsc{cft}s (one free and one interacting),  
the interacting one having a central charge proportional to $Q_1$, the number of fundamental strings 
that build the AdS$_3$ background. The long strings emission by the D0-brane decay decreases their number. 
We expect eventually that after non-perturbative effects are taken into account, 
in order that the energy released into closed strings radiation is of the order of the D0-brane mass, 
$\nicefrac{\delta c}{c} = \nicefrac{\delta Q_1}{Q_1}$ will be of order $\nicefrac{1}{\sqrt{Q_1}}$. 
If the cutoff that we use turns out to be appropriate, the relative variation of the central charge 
is very small in the perturbative regime.

This paper is organized as follows. 
In section~\ref{secbraneads}  we recall some 
aspects of string theory in AdS$_3$ and discuss  
the non-\textsc{bps} D-particle in AdS$_3 \times$ S$^3 \times$ T$^4$. In
section~\ref{secrolbound} we briefly review  the rolling tachyon  
in flat space-time and construct the rolling tachyon boundary state in AdS$_3$. 
Section~\ref{secemission} is devoted to the analysis of closed string 
emission in AdS$_3 \times$S$^3 \times$ T$^4$. This is the main result of this work. 
We discuss also the emission of open strings. 
In section~\ref{secholog} we make some comments on the holographic 
interpretation of the decay.

\boldmath
\section{D-particles and D-instantons in AdS$_3 \times$ S$^3 \times$ T$^4$}
\unboldmath
\label{secbraneads}
In this section we construct in detail the unstable D-particle in 
AdS$_3 \times$ S$^3 \times$ T$^4$. The reader interested mostly by 
the decay of the brane can move directly to section~\ref{secrolbound}.

We will first set the stage with a short review of 
string theory on AdS$_3$. Then we will recall 
the construction of the D(-1)- and D0-branes in AdS$_3$ 
at the microscopic boundary \textsc{cft} level 
and  embed the latter in the AdS$_3 \times$ S$^3 \times$
T$^4$ type \textsc{iib} superstring background, the example we 
will follow in the next sections.  

\boldmath
\subsection{Strings and D-branes in AdS$_3$}
\unboldmath
String theory on AdS$_3$, with an \textsc{ns-ns} 2-form flux, 
is described by the \textsc{wzw} model for (the universal cover of)
$\slr$. In the following we will consider the supersymmetric 
\textsc{wzw} model that is used to construct superstring 
theories. The corresponding background fields  for  
a supersymmetric affine $\widehat{\mathfrak{sl}}(2,\mathbb{R})$ algebra\footnote{
Containing a purely bosonic $\widehat{\mathfrak{sl}}(2,\mathbb{R})$ algebra at level $k+2$.} at level $k$ are:
\begin{eqnarray}
\di s^2 &=& \alpha ' k \left[ \di \rho^2 + \sinh^2 \rho \, \di \phi^2 
-\cosh^2 \rho\,  \di t^2 \right]\nonumber \\
H  & = & 2 \alpha' k \cosh \rho \sinh \rho \ \di \rho \wedge 
\di \phi \wedge \di t \, ,
\label{globalcord}
\end{eqnarray}
using the global coordinates. The dilaton field is a constant that we choose such that the 
string coupling $g_s=\exp \Phi$ is small, in order 
to justify worldsheet \textsc{cft} techniques. The central charge 
of this superconformal theory is $c = 9/2+6/k$. 

The closed string spectrum splits into standard 
$\widehat{\mathfrak{sl}}(2,\mathbb{R})$ representations ($w =0$) and 
twisted ones ($w \neq 0$) obtained by an outer automorphism called 
{\it spectral flow}~\cite{Henningson:1991jc,Maldacena:2000hw}. 
The weights of the primaries read, in the \textsc{ns} sector:
\begin{equation}
L_0 = -\frac{j(j-1)}{k} -w m + \frac{k}{4}w^2 \quad , \qquad 
\bar L_0 = -\frac{j(j-1)}{k} -w \bar m + \frac{k}{4}w^2 \ , 
\label{dimprim}
\end{equation}
where $(m,\bar m)$ label primaries of the $\hat{\mathfrak{u}}_1$ 
elliptic compact sub-algebra $(J^3,\bar J^3$). Space-time energy is given 
by $E= m + \bar m$ whereas the angular momentum (conjugate to $\phi$) 
is $n= m-\bar m$. The $\mathfrak{sl}(2,\mathbb{R})$ representations 
appearing in the unitary closed string spectrum fall 
into two classes. The {\it discrete representations} with~\cite{Maldacena:2000hw} 
\begin{equation}
\frac{1}{2} < j< \frac{k+1}{2}\, 
\label{spinbound}
\end{equation} 
are related to string worldsheets trapped inside AdS$_3$, corresponding
to classical solutions obtained by spectral flow of time-like geodesics. 
The {\it continuous representations} with 
$j = \frac{1}{2} + i P$, $P \in \mathbb{R}_+$ 
are related to closed strings whose wave-functions are 
delta-function normalizable. They correspond to circular  
long string solutions growing linearly w.r.t. global time, 
with asymptotic winding number $w$, obtained by the spectral flow of 
space-like geodesics.

It is convenient for many computations to decompose the \textsc{wzw} model 
in terms of the coset $\slc$, 
the Euclidean 2D black hole~\cite{ Elitzur:1991cb,Mandal:1991tz,Witten:1991yr}, 
and a time-like boson: 
\begin{equation}
\slr_k \sim  \frac{\slc|_k \times \text{U(1)}_{-k}}{\zi} \, ,
\label{decompslr}
\end{equation} 
We refer the reader to appendix~\ref{appeads} and to~\cite{Israel:2005ek} for more algebraic details.

\boldmath
\paragraph{Localized D-branes in AdS$_3$}
\unboldmath
As in any \textsc{wzw} model, the {\it symmetric} D-branes 
in $\slr$ are defined by the gluing conditions for the currents of the affine
algebra, which include a possible twist by an (outer) automorphism
of the algebra preserving the gluing conditions for the
($\cn=1$ super-)conformal algebra~\cite{Alekseev:1998mc}. 

The D-instanton that we will be interested in is given by the 
trivial gluing conditions (in the open string channel)~\cite{Bachas:2000fr}:
\begin{equation}
J^3 (z) = \bar J^3 (\bar z) |_{z=\bar z} \quad , 
\qquad J^{\pm} (z) =  \bar{J}^{\pm} (\bar z) |_{z=\bar z} \, .
\label{glueinst}
\end{equation}
It corresponds to a point-like object in AdS$_3$ space-time, localized 
at $\rho=t=0$ in global coordinates. The exact construction of the 
boundary state in Lorentzian AdS$_3$~\cite{Israel:2005ek} follows
from the same logic that we discussed above, starting with the D-branes in 
$\slc$~\cite{Ribault:2003ss,Eguchi:2003ik,Israel:2004jt,Fotopoulos:2004ut} that 
can be obtained either from the gauging of H$_3^+$~\cite{Ponsot:2001gt}, the Euclidean 
continuation of AdS$_3$ or using the conformal 
bootstrap of the $\cn =2$ 
superconformal algebra with
$c>3$~\cite{Ahn:2003tt,Hosomichi:2004ph}. The one-point function for an \textsc{ns-ns} primary 
field $V^{j}_{m \bar m w} (z,\bar z)$ in the presence of   
this localized "S-brane" reads~\cite{Israel:2005ek}: 
\begin{equation}
\langle V^{j}_{m \bar m w} (z,\bar z) \rangle = \frac{1}{|z-\bar
  z|^{2\Delta_{jmw}} }\ 
 \frac{2i\pi}{(2k)^{3/4}}\, \nu_k^{\nicefrac{1}{2}-j}  
 \  \frac{\Gamma (j+m-\frac{kw}{2}) 
\Gamma (j- m+\frac{kw}{2})}{\Gamma (2j-1) \Gamma (1+ \frac{2j-1}{k})}\
\delta_{m-\bar m,0} \, , 
\label{dinstonept}
\end{equation}
where $\nu_k = \Gamma (1-\nicefrac{1}{k})/\Gamma(1+\nicefrac{1}{k})$ and 
$\Delta_{jmw}$ is the conformal dimension of the primary as given by 
eqn.~(\ref{dimprim}). It contains couplings to closed strings both in the continuous and discrete 
representations. The former are obtained by evaluating~(\ref{dinstonept}) for 
$j \in \nicefrac{1}{2} + i \mathbb{R}_+$, while the latter are the {\it residues} at the poles 
of the one-point function  for real $j$. 
In the semi-classical regime $k\to \infty$, these couplings can be found by
applying the distribution $\delta (\rho) \delta (t)$ to 
the (delta-function) normalizable wave-functions on AdS$_3$, 
as in~\cite{Ribault:2003ss} for the coset model. 

The open string annulus amplitude is obtained by 
closed/open channel duality using this one-point function\footnote{Details 
about this standard \textsc{bcft} computation can be found e.g. in the 
lecture notes~\cite{Schomerus:2002dc}, section 3.} 
Still concentrating only on the \textsc{ns} sector, we get 
in Lorentzian space-time 
the {\it identity representation} of $\slr$\footnote{
On the single cover of $\slr$ the open string spectrum contains also 
spectral flowed representations associated with 
open strings winding around the periodic time direction. This observation 
allows to check that on the universal cover of $\slr$, the boundary state corresponds to 
only one copy of the D-instanton and not an array of them along the
Euclidean time direction. 
} (with $q= \exp 2i\pi \tau$):\footnote{The fermionic characters 
are written using the theta-function $\vartheta \oao{a}{b} (\tau ) = \sum_{n \in \zi} 
q^{\frac{1}{2}(n+a/2)^2} e^{i\pi (n+a/2)b}$ where $a,b \in \zi_2$ label the different spin structures. 
$a=0$ (resp. $a=1$) corresponds to the Neveu-Schwarz (resp. Ramond) sector. The $b=1$ sectors 
have a $(-)^F$ insertion in the trace.}
\begin{equation}
Z (\tau ) = 
\sum_{r \in \zi} ch_\mathbb{I} \oao{0}{0} (r;\tau) 
\frac{q^{-\frac{r^2}{k}}}{\eta (\tau )} 
\left( \frac{\vartheta \oao{0}{0} (\tau)}{\eta (\tau)}\right)^{1/2}\, ,
\label{openpartinst}
\end{equation}
that we decomposed in terms of U(1) characters and the ``string functions'' 
$ch_\mathbb{I} \oao{0}{0} (r;\tau)$ of the super-coset $\slc$  
for the  identity representation. To get a proper interpretation of this 
brane as a D-instanton one should consider an Euclidean target space instead, 
i.e. Wick rotate AdS$_3$ to H$_3^+$. We would like also to stress that, 
unlike in the closed string sector, the identity representation --~containing 
in particular the vacuum state~-- is present in this sector 
of open strings attached to a localized brane. 

Using the coset decomposition~(\ref{decompslr}) one can 
construct {\it symmetry-breaking D-branes}~\cite{Maldacena:2001ky} 
associated with the same boundary 
conditions in the $\slc$ coset. It boils down to replacing Dirichlet boundary conditions 
by Neumann ones for the time-like boson, changing 
appropriately the overall normalization of the boundary state 
to satisfy the Cardy condition (i.e. such that the identity appears in the open 
string spectrum with coefficient one).  This D-brane is 
interpreted geometrically as a D-particle sitting at the origin 
of AdS$_3$~\cite{Quella:2002ns}.
The associated one-point function for an \textsc{ns-ns} primary is~\cite{Israel:2005ek}:\footnote{In the 
following we will not write explicitly the $(z,\bar z)$ dependence of
the one-point functions any more.}
\begin{equation}
\langle V^{j}_{m \bar m w} \rangle = i \pi \left(\frac{2}{k}\right)^{1/4}
\nu_k^{\nicefrac{1}{2}-j} 
 \  \frac{\Gamma (j+\frac{kw}{2}) 
\Gamma (j-\frac{kw}{2})}{\Gamma (2j-1) \Gamma (1+ \frac{2j-1}{k})}\
\delta_{m-\bar m,0}\ \delta (m+\bar m) \, .
\label{d0onepf}
\end{equation}
The Kronecker delta $\delta_{m-\bar m,0}$ comes from the boundary conditions in $\slc$, while 
the Dirac delta-function $\delta (m+\bar m)$ corresponds to the Neumann boundary conditions 
for the free boson (and is related to the time-translation symmetry preserved by the brane). 
The only solution of these antagonistic constraints is $m=\bar m = 0$. 
Using again closed/open channel duality it gives the open string partition function as follows:
\begin{equation}
Z  = V_1 \sum_{r \in \zi} ch_\mathbb{I}\oao{0}{0} (r;\tau) \int \frac{\di E}{2\pi} \
\frac{q^\frac{-E^2}{k}}{\eta} \left( 
\frac{\vartheta \oao{0}{0} (\tau)}{\eta (\tau)}\right)^{1/2} \, ,
\end{equation}
where $V_1 \equiv 2\pi \delta (0)$ is the "volume" of the time direction. 
This open string spectrum looks similar to what we would obtain 
in a background containing $\slc \times \mathbb{R}^{0,1}$, for instance 
a non-critical superstring (the divergence coming from the integration 
over space-time energy $E$ has to be regularized as in this 
flat space-time). We will see below that 
for the rolling tachyon \textsc{bcft} describing its decay 
the $\slc$ and time-like $U(1)$ open string sectors  
are not decoupled anymore, since closed string primaries with $m\neq0$ contribute 
to the annulus amplitude. 


\boldmath
\subsection{The non-BPS D-particle in AdS$_3 \times$ S$^3 \times$ T$^4$}
\unboldmath
We consider now the full type \textsc{iib} superstring theory on 
AdS$_3 \times$ S$^3 \times$ T$^4$ with \textsc{ns-ns} fluxes.
It can be obtained as the near-horizon geometry for a collection of 
$k$ coincident NS5-branes and $Q_1$ fundamental strings smeared on the 
four-torus~\cite{Boonstra:1997dy}. The level of the $\slr$ algebra is $k$ 
(corresponding to anti-de Sitter space-time of radius $\sqrt{\alpha' k}$), while the six-dimensional string 
coupling constant is fixed in the near-horizon limit to 
\begin{equation} g_6 = \frac{g_s}{\sqrt{v_4}}=\sqrt{\frac{k}{Q_1}}\, ,
\end{equation}
where $v_4$ is the T$^4$ volume in strings units. 
We will take this solution as a concrete example to embed the 
non-\textsc{bps} D-particle, and later the associated rolling tachyon,  
in a superstring background.

Besides the AdS$_3$ factor, the three-sphere with 
\textsc{ns-ns} flux is described as an SU(2) super-\textsc{wzw} model 
at level $k$. An important point is that, since this curved background 
is made only of \textsc{wzw} models, the worldsheet fermions are free even though 
the background is non-trivial. 
The space-time supercharges are then constructed exactly as in flat space 
using the associated spin fields~\cite{Friedan:1985ge}, with however the extra condition 
$\gamma^{012345}\zeta_\textsc{l,r} = \zeta_\textsc{l,r}$ on the two ten-dimensional spinors, giving 
overall 16 supercharges~\cite{Giveon:1998ns}. 

We wish now to consider D-branes in this superstring theory, 
more specifically to embed the AdS$_3$ D0-brane discussed 
above. For the SU(2) part we choose the ``elementary'' boundary state 
associated to the identity representation, i.e. the S$^2$ brane of minimal 
volume that is viewed as a point on S$^3$ in the semi-classical limit. Other choices, 
corresponding to ``large'' two spheres, can be viewed as bound states 
of these elementary branes~\cite{Alekseev:2000fd}. In order to simplify 
the notation we take an orthogonal torus. We denote by $R_i$, $i=1,\ldots ,4$ 
the radii of the circles. 

There are two reasons for the D0-brane not to be 
\textsc{bps}. First, a D0-brane in type \textsc{iib} has the wrong dimensionality 
to be supersymmetric, as it doesn't couple to the \textsc{r-r} sector.  
Second, even in type \textsc{iia}, this brane would be non-supersymmetric 
because it breaks the $\slr$ symmetry which prevents from decoupling 
the worldsheet fermions and writing the space-time supercharges with 
spin fields, as was done in~\cite{Giveon:1998ns} for the closed string 
sector. It is also not possible to use a Gepner-like construction~\cite{Gepner:1987qi} 
because the R-charges of the $\cn = 2$ worldsheet superconformal algebra 
associated with the $\slc$ coset are not integral. One can check that this D0-brane 
of type \textsc{iia}, although non-\textsc{bps}, is stable (i.e. the 
open string sector does not contain tachyons). 

Let's come back to the non-\textsc{bps} 
D0-particle in type \textsc{iib} superstrings on AdS$_3 \times$ S$^3 \times$ T$^4$. 
Its open string partition function reads~: 
\begin{equation}
Z (\tau )=  \sum_{a \in \zi_2}(-)^a   \sum_{\{ w^i\}\in \zi^4 } 
\frac{q^{\frac{(R_i w^i)^2}{\alpha'}}}{\eta^3 (\tau )}
\sum_{r \in \zi} ch_\mathbb{I} \oao{a}{0} (r;\tau) 
\int \frac{\di E}{2\pi} \, q^{-\frac{E^2}{k}} \ 
\chi^{0} (\tau )\ \frac{\vartheta \oao{a}{0}^3 (\tau )}{\eta^3 (\tau )} \, . 
\label{dpartannulus}
\end{equation}
where $\chi^{0} (\tau )$ is the $\widehat{\mathfrak{su}(2)}$ character of the trivial
representation. The \textsc{ns} and \textsc{r} sectors correspond  
to $a=0$ and $a=1$ respectively. 
Let us now study open string tachyonic modes on this D-particle,  
since we will be ultimately interested in finding the time-dependent 
solutions associated with rolling down these negative directions of the potential 
in the open string field theory. By taking the identity in the trivial 
representation for the $\slc$ factor ($r=0$) as well as for SU(2) we find 
a tachyon similar to the one that appears on a non-\textsc{bps} D-brane 
in flat space-time, with mass squared 
$m^2_\textsc{tach} = -\nicefrac{1}{2 \alpha '}$ (using flat space-time normalization).
If we choose the D-particle to be point-like on the three-sphere, there are no 
other tachyonic modes provided the compactification torus is large enough. 
For branes wrapping a non-vanishing $S^2$, associated with an SU(2)
representation of spin $\hat \jmath$, there are extra negative 
modes of mass  squared $m^2_\textsc{tach} = -\nicefrac{1}{2 \alpha '}+\nicefrac{\tilde 
\jmath (\tilde \jmath +1)}{\alpha'k}$ if $\tilde \jmath \in \{ 1,2,\ldots, 2\hat
\jmath$ \} is small enough. They correspond to inhomogeneous decay on their two-sphere 
worldvolume. However the rolling tachyon \textsc{bcft} is  more complicated 
since the boundary deformation involves a non-trivial boundary operator in the 
$SU(2)$ sector. In the following we will concentrate on point-like 
D-particles with $\hat \jmath = 0$.

\boldmath
\section{The rolling tachyon boundary state in AdS$_3$ }
\unboldmath
\label{secrolbound}
Having constructed the non-\textsc{bps} D-brane of interest in 
the  AdS$_3 \times$ S$^3 \times$ T$^4$ type \textsc{iib} superstring
background, and having found its tachyonic modes, we will now study
the boundary state corresponding to the decay of this D-particle towards the minimum 
of the tachyon potential, the closed string vacuum. We will first review some aspects of 
the rolling tachyon boundary conformal field theory in flat space-time, then 
apply these techniques to the anti-de Sitter case.

\subsection{Rolling tachyon in flat space-time: short review}
\label{rollflat}
The rolling tachyon  is an exact solution of the boundary 
worldsheet \textsc{cft} describing the decay of an unstable 
D-brane~\cite{Sen:2002nu}. For simplicity we discuss first the bosonic string.  
It comes in two versions corresponding to the boundary marginal deformations:
\begin{subequations}
\begin{align}
\delta S_\textsc{full} &= \lambda \oint_{\p \Sigma} \di \ell \ \cosh \left(
\nicefrac{X^0(\ell)}{\sqrt{\alpha '}}\right) \label{rolbounddef}\\
\delta S_\textsc{half} &= \frac{\tilde{\lambda}}{2} \oint_{\p \Sigma} \di \ell \ 
e^{\nicefrac{X^0(\ell )}{\sqrt{\alpha '}}}
\label{halfrolbounddef}
\end{align}
\end{subequations}
The first one describes a time-dependent process in which incoming radiation
conspires --~involving a considerable fine-tuning~-- to create a 
non-\textsc{bps} D-brane at $x^0 = 0$ that will eventually  decay back to the
closed string vacuum. The second one corresponds to the 
decay of a non-\textsc{bps} D-brane prepared at past 
infinity $x^0 \to - \infty$. The parameter $\tilde \lambda$ is not 
important in this case since it can be absorbed by a time
translation. In both cases the decay is homogeneous along the D-brane
longitudinal directions.

As always in string theory, in order to study this worldsheet \textsc{cft} corresponding 
to a time-dependent process in spacetime we need to use an
analytic continuation that is not uniquely defined. It turns out in this case
that the rolling tachyon \textsc{bcft} can be obtained from two rather different types of Euclidean models. 
The first one, which is more naturally related to the ``full S-brane'' 
solution~(\ref{rolbounddef}), is the boundary sine-Gordon 
theory~\cite{Callan:1994ub,Polchinski:1994my,Recknagel:1998ih,Gaberdiel:2001zq}. 
It consists in a $c=1$ free theory of a space-like boson, with the boundary 
deformation $\lambda \oint_{\p \Sigma} \di \ell \ \cos
\nicefrac{X(\ell)}{\sqrt{\alpha '}}$. 
The construction of the boundary state uses as a basis of Ishibashi 
states the higher order Virasoro primaries that occur 
for dimensions $\Delta \in (\zi)^2/4$ and the underlying $\widehat{\mathfrak{su}}(2)_1$ 
symmetry of the model. The Lorentzian theory does not have such primaries (because all 
the Virasoro representations except the identity are non-degenerate) and such a symmetry, 
therefore the Wick rotation is quite intricate.  Some 
components of the boundary state obtained this way grow  exponentially with 
time~\cite{Okuda:2002yd} but do not correspond to on-shell physical 
closed string states. Even though they 
have been interpreted as conserved charges in the  
of two-dimensional strings context~\cite{Sen:2004zm} their interpretation 
in critical strings and their physical relevance are unclear.

The other route to the rolling tachyon is a particular limit 
of Liouville theory~\cite{Gutperle:2003xf} involving two analytic 
continuations. Liouville theory has a central 
charge $c=1+6(b+b^{-1})^2 >25$ and to get a $c=1$ theory 
one should consider the analytic continuation $b\to i$. In this 
limit the Liouville potential ceases to be a wall and becomes periodic.\footnote{
As shown in~\cite{Schomerus:2003vv} this unitary \textsc{cft} is identical to the $c\to 1$ limit 
of Virasoro minimal models studied in~\cite{Runkel:2001ng}.}
 For the boundary \textsc{cft} that we consider, namely 
Liouville theory with boundary conditions corresponding to an extended 
D-brane (called \textsc{fzzt} brane~\cite{Fateev:2000ik,Teschner:2000md})
we take the limit of vanishing Liouville potential, while keeping fixed 
the boundary cosmological constant $\tilde \lambda$. 
The other analytic continuation corresponds to Wick-rotating the Liouville field 
$\phi \to iX^0$ to get a time-like boson;  it gives then 
the ``half-S-brane'' solution~(\ref{halfrolbounddef}). However the 
\textsc{bcft} data, excepting the one-point function, is not 
analytic in the Liouville momentum for 
$b \in i\mathbb{R}$~\cite{Schomerus:2003vv,Fredenhagen:2004cj}.

The one-point function for the \textsc{fzzt}-brane  in time-like Liouville
theory with $c=1$ and vanishing bulk cosmological constant 
reads~\cite{Gutperle:2003xf}:
\begin{equation}
\langle e^{i k_0 X^0} \rangle =  \frac{1}{\sqrt{2\pi}}
\left(\pi \tilde \lambda \right)^{-i\sqrt{\alpha '} k_0}\ 
\frac{\pi}{ \sinh 
\pi \sqrt{\alpha'} k_0  }\, .
\label{polesrolltach}
\end{equation}
This is what we will consider in the following as defining the coupling
between the rolling tachyon~(\ref{halfrolbounddef}) and closed
strings.\footnote{There 
is a complication arising for $k_0=0$, which is a simple pole 
of the one-point function, related to the fact that  
the identity representation is degenerate. It splits into infinitely 
many Virasoro ireps labeled by $J \in \mathbb{Z}_{\geqslant 0}$, whose highest 
weight state $| J,0,0 \rangle$, of dimension $J^2$, is a polynomial in $X^0$ derivatives 
$\mathcal{O} (\p X^0, \p^2 X^0, \pb X^0, \pb^2 X^0, \cdots)$. 
A more correct definition of the boundary state would be 
$|B\rangle = \sqrt{\pi/2} \, \mathcal{P}\! \int \di k_0 \, 
(\pi \tilde{\lambda})^{-i\sqrt{\alpha '} k_0}\sinh^{-1} (\pi \sqrt{\alpha'}
k_0) | k_0 \rrangle +\sum_{J=0}^{\infty} |J,0,0 \rrangle $.}

The extension of the above results to the superstring case is rather
straightforward, applying the same manipulations to $\cn =1$
super-Liouville theory with a boundary. 
The boundary deformation associated with the rolling tachyon now reads~\cite{Larsen:2002wc}:
\begin{equation}
\delta S = \frac{\tilde \lambda}{2} 
\oint_{\p \Sigma} \di \ell  \ \psi^0 e^{X^0/\sqrt{2\alpha'}} 
\sigma^1
\label{defsuper}
\end{equation} 
where $\sigma^1$ is a Chan-Patton factor coming from  
a fermionic zero mode on the worldsheet boundary. It is studied by  
analytic continuation of the super-\textsc{fzzt} D-brane~\cite{Ahn:2002ev,Fukuda:2002bv}. 
Then the one-point function 
for an \textsc{ns-ns} primary in the presence of this boundary deformation  
is similar to the bosonic strings result~(\ref{polesrolltach}):~\footnote{
See~\cite{Shelton:2004ij} for another derivation of this result using free field 
correlators.}
\begin{equation}
\langle e^{i k_0 X^0} \rangle =  \frac{1}{\sqrt{2\pi}}
\left(\frac{\pi \tilde{\lambda}}{2}\right)^{-i\sqrt{2\alpha '} k_0} 
\frac{\pi}{ \sinh 
\pi \sqrt{\frac{\alpha'}{2}} k_0  }\, .
\label{NSNSrolltach}
\end{equation}
 
Operators in the \textsc{r-r} sector of the closed superstring theory  
get non-zero one-point functions   
once the rolling tachyon perturbation is turned on. More explicitly, 
we consider the spin fields $\sigma^{\epsilon}$, with $\epsilon = \pm 1$, creating 
the two Ramond ground states for $\psi^0$, the worldsheet fermionic super-partner of $X^0$. 
The one-point function in the \textsc{r-r} sector depends on the two possible gluing 
conditions for the $\cn=1$ supercurrent on the real axis: 
\begin{equation}
G (z) = \xi \, \tilde{G} (\bar z)|_{z=\bar z} \quad , \qquad \xi = \pm 1 \, . 
\end{equation}
For both signs, the one-point function  
for the super-\textsc{fzzt}-brane  in time-like super-Liouville
theory with $c=3/2$ and vanishing bulk cosmological constant
reads:
\begin{equation}
\langle e^{i k_0 X^0(z,\bar z)} \sigma^\epsilon (z) 
\tilde{\sigma}^{\bar \epsilon} (\bar z)  \rangle =  \frac{\epsilon^{\frac{\xi+1}{2}} }{\sqrt{2\pi}} \  
\left(\frac{\pi \tilde{\lambda}}{2}\right)^{-i\sqrt{2\alpha '} k_0}
\frac{\pi}{\cosh 
\pi \sqrt{\frac{\alpha'}{2}} k_0 }\, \delta_{\epsilon,\bar \epsilon} \, .
\label{rolltachrr}
\end{equation}
Recall that a non-\textsc{bps} 
D-brane with constant open string tachyon profile does not couple to the \textsc{r-r} fields. 
In the effective action the couplings between the tachyon field and the \textsc{r-r} forms 
are derivative.

\boldmath
\subsection{D0-brane decay in AdS$_3$}
\unboldmath
\label{rollads}
We have now prepared the ground 
to construct the boundary state of the rolling tachyon in 
AdS$_3$. We follow the approach based on time-like Liouville theory that we reviewed above.

Let us first recall that we showed in the previous section that 
the unstable D0-brane in type \textsc{iib} superstring theory 
on AdS$_3 \times$ S$_3 \times$T$^4$ contains on its worldvolume an open string 
tachyon built with the {\it identity} of the $\slc$ super-coset. Therefore 
one can turn on the same boundary deformation~(\ref{defsuper}) that was
discussed above in flat-space-time, built with the identity of 
the $\slc$ conformal field theory. According to the general analysis 
of boundary deformations performed in~\cite{Recknagel:1998ih}, this 
deformation of the AdS$_3$ superstring is exactly marginal,  therefore it
gives an exact rolling tachyon solution of the open string field theory. 
Moreover we shall see below that the one-point function giving the coupling
between closed string modes and the rolling tachyon can be computed exactly. 
To simplify the equations we will consider till the end of this section only 
the $\slr$ part of the one-point functions. In the full 
\mbox{AdS$_3 \times$ S$_3 \times$T$^4$} 
background the one-point function will be the product of the $\slr$ piece  
with the SU(2)$\times$U(1)$^4$ piece which is standard, together with the appropriate \textsc{gso} 
projection.

As reviewed in more detail in appendix~\ref{appeads}, 
the primary states in the  Minkowskian $\slr$ theory 
--~characterized by their $\slr$ spin $j$, the eigenvalues 
$(m,\bar m)$ of the elliptic sub-algebras 
$J^3_0$ and $\bar J^3_0$ and the spectral 
flow sector $w$~-- are defined 
as primary states in the ``T-dual'' theory $\slc \times \mathbb{R}^{0,1}$. 
For \textsc{ns-ns} primaries we have the decomposition:  
\begin{equation}
V^{j\, , \mathfrak{sl}_2}_{m \bar m w}  = 
V^{j\, , \mathfrak{sl}_2/\mathfrak{u}_1}_{m-\frac{kw}{2} \, ; \ -\bar m +
  \frac{kw}{2}}\ 
e^{i\sqrt{\frac{2}{k}} \left[ m X^0 (z) + \bar m \tilde X^0 (\bar z)
  \right]} 
\label{primdef}
\end{equation}
where $X^0 (z,\bar z)$ is a canonically normalized free time-like boson 
(with $\alpha ' = 2$). To deal with the \textsc{r-r} sector it is convenient 
to bosonize the fermionic superpartners of the 
currents $J^{\pm}$ and $\bar J^{\pm}$ as 
$2 i\p H_1 =\psi^+ \psi^-$ and $2 i\pb \tilde H_1 = \tilde \psi^+ \tilde
\psi^-$. Then we have the following decomposition:
\begin{equation}
V^{j\, , \mathfrak{sl}_2}_{m \bar m w} e^{\frac{i}{2} 
(s H_1 + \bar s \tilde H_1)} \sigma^{\epsilon, \bar \epsilon } = 
V^{j\, , \mathfrak{sl}_2/\mathfrak{u}_1 \, (s, \bar s)}_{m-\frac{kw}{2} 
\, ; \ - \bar m +  \frac{kw}{2}}\ \ 
e^{i\sqrt{\frac{2}{k}} \left[ m X^0 (z) + \bar m \tilde X^0 (\bar z)
  \right]}   \sigma^{\epsilon, \bar \epsilon } \, .
\label{primdefrr}
\end{equation}
with $s,\bar s = \pm 1$. 
The one-point function for a D-particle, in the presence of the boundary
deformation~(\ref{defsuper}), can be obtained as follows. First, due 
to the decomposition~(\ref{primdef}) of $\slr$ vertex operators into an 
$\slc$ piece and a free boson piece, 
the contribution to the rolling tachyon  one-point function  
from the $\slc$ vertex operator will be the same 
as its contribution to the D(-1)-brane,  
see eqn.~(\ref{dinstonept}), up to the overall normalization factor. 
Indeed the boundary deformation~(\ref{defsuper}) corresponding to the rolling tachyon 
is built with the {\it identity} operator 
of $\slc$ \textsc{cft} and therefore cannot 
change the $\slc$ contribution to the coefficients of the 
boundary state.

Second, the contribution from the time-like boson is identical, for 
a given value of the spacetime energy, to the one-point function for the  
rolling tachyon  in flat space-time~(\ref{polesrolltach}). Indeed this is 
a free boson, in the sense that, 
expanding an $\widehat{\mathfrak{sl}}(2,\mathbb{R})$ highest 
weight module in terms of the horizontal Cartan subalgebra $J_0^3$, 
for a given value of $m$ the sub-module splits into an
$\widehat{\mathfrak{sl}}(2,\mathbb{R})/\widehat{\mathfrak{u}}_1$ 
highest weight module and a free $\widehat{\mathfrak{u}}_1$ one. Moreover the $\slc$ 
contribution to the one-point function is analytic in  $m$ 
so one can choose imaginary values corresponding to ``resonances'' 
for which the Liouville couplings are given by free field computations.

Assembling everything together we get the one-point 
function in the presence of the decaying D-particle in
AdS$_3$. It leaves a freedom for the overall normalization 
of the one-point function, that we fixed by an analogue 
of the Cardy condition.\footnote{One requires that the annulus amplitude in the 
open string channel has a normalization compatible with 
the regularized density of open string states~\cite{Teschner:2000md}.} 
Explicitly, we get 
first the one-point function for an \textsc{ns-ns} primary as follows: 
\begin{equation}
\langle V^{j\, , \mathfrak{sl}_2}_{\nicefrac{E}{2}\ \nicefrac{E}{2}\ w} \rangle = 
\frac{\pi}{(2k)^{\nicefrac{3}{4}}}  \ \nu_k^{\nicefrac{1}{2}-j}
 \  \frac{\Gamma \left( j+\frac{kw - E}{2} \right) 
\Gamma \left(j-\frac{kw - E}{2}\right)}{
\Gamma (2j-1) \Gamma (1+ \frac{2j-1}{k})}\
\frac{ \left( \frac{\pi \tilde{\lambda}}{2}\right)^{-i\sqrt{\frac{2}{k}}E}}{\sinh 
\frac{\pi E}{\sqrt{2k}}} \, .
\label{adstachonept}
\end{equation}
This is one of the main results of the paper, that we will use 
in the following to study the physics of the decay. Compared 
to the static D0-brane one-point function, eqn.~(\ref{d0onepf}), 
there is no constraint $\delta (E)$ because the brane 
is time-dependent. Single closed string states with angular 
momentum (i.e. $m - \bar m \neq 0$) don't couple to the D0-brane, 
because the latter carries no angular momentum in its rest frame.

One can find the profile of the closed string wave-function for the brane  
by inverse Fourier transform 
in the sector $w=0$, using a basis of (delta-function)-normalizable 
functions on the $\slr$ group manifold, i.e. matrix elements in discrete and continuous representations. 
The one-point function (\ref{adstachonept}) is the product 
of the one-point  functions for a D(-1) instanton and for the rolling tachyon in flat
space-time; therefore the profile  is obtained by the convoluted product of 
these two. In the semi-classical limit, i.e. neglecting the factor 
$\Gamma (1+\nicefrac{2j-1}{k})$ and removing the upper 
bound~(\ref{spinbound}) on $j$, the Fourier transform of the $D(-1)$
coupling is proportional to $\delta (\rho) \delta (t)$. We find that 
the spacetime profile of the rolling tachyon in the semi-classical limit is  
\begin{equation} 
\Psi (\rho,t,\phi) \propto \frac{\delta (\rho)}{1+(\nicefrac{\pi \tilde \lambda}{2})^2 e^{\sqrt{2k}\, t}}.
\end{equation} 
It describes the decay of an unstable D-particle sitting at the center of
AdS$_3$, prepared at past infinity $t\to - \infty$ in global time. 
We will use later the same one-point function to compute the emission of closed 
strings coupling to this time-dependent brane.

Let us now move to the \textsc{r-r} sector. With the 
conventions set above, see eqn.~(\ref{primdefrr}),  
the one-point function for a \textsc{r-r} primary 
of $\slr$ in the presence of the rolling tachyon reads:
\begin{multline}
\langle
V^{j\, , \mathfrak{sl}_2}_{\nicefrac{E}{2}\, \nicefrac{E}{2}\, w} e^{\frac{i}{2} 
(s H_1 + \bar s \tilde H_1)} \sigma^{\epsilon, \bar \epsilon}
\rangle =  \epsilon^{\frac{\xi+1}{2}} e^{-\frac{i\pi s \hat s}{2}} 
\frac{\pi}{(2k)^{\nicefrac{3}{4}}}  \ \nu_k^{\nicefrac{1}{2}-j}
\ \times \\ \times \ 
 \  \frac{\Gamma \left( j+\frac{kw - E+s}{2} \right) 
\Gamma \left(j-\frac{kw -E - \bar s }{2}\right)}{
\Gamma (2j-1) \Gamma (1+ \frac{2j-1}{k})}
\frac{\left( \frac{\pi \tilde{\lambda}}{2}\right) ^{-i\sqrt{\frac{2}{k}}E}}{\cosh
\frac{\pi E}{\sqrt{2k}}} \  \delta_{s, -\bar{s}} 
\delta_{\epsilon, \bar \epsilon}  
\label{adstachoneptrr}
\end{multline}
with a parameter $\hat s \in \zi_4$ such that  
$\hat s$ is even for $\xi = 1$ and odd otherwise. 
In the full supersymmetric type \textsc{iib} background the boundary state 
is  a linear combination of the $\xi = 1$ and $\xi = -1$ boundary 
states in the \textsc{ns-ns} and \textsc{r-r} sectors in the usual way compatible 
with the closed string \textsc{gso} projection~\cite{Gaberdiel:2000jr}.

We recall that this one-point function contains both couplings 
to the discrete representations (the residues of the simple poles on the 
real axis in the $j$-plane) and couplings to the continuous 
representations with $j = \frac{1}{2}+ iP$. 
Therefore one of the most prominent features of the 
the decay of this D-brane is that it produces 
long strings that ``escape'' to space-like infinity. 
In the following section we will make this more 
precise and compute the emission of closed strings as the imaginary 
part of the annulus amplitude.

\paragraph{Conserved energy}
The conserved energy of the rolling tachyon can be found from the 
``conservation law'' for the boundary state:
\begin{equation}
(Q_B + \bar Q_B ) |B\rangle  = 0\, , 
\label{brstbdy}
\end{equation}
in order to satisfy the closed string field theory equations of motion;  
$Q_B$ and $\bar Q_B$ are the left and right contribution to the 
\textsc{brst} charge. We refer the reader to the review~\cite{Sen:2004nf} for 
more details and references. The expansion of the boundary state for the 
rolling tachyon in an $AdS_3 \times \mathcal{M}$ bosonic string background at level $k$ 
contains the terms
\begin{multline}
| B \rangle \propto 
  \int \ 
\di j\ \   \nu_{k-2}^{\nicefrac{1}{2}-j}  \sum_{w} \int \di E \ 
 \frac{\Gamma \left( j+\frac{kw - E}{2} \right) 
\Gamma \left(j-\frac{kw -E }{2}\right)}{
\Gamma (2j-1) \Gamma (1+ \frac{2j-1}{k-2})}  \\   
\left( \frac{\pi(\pi {\lambda})^{-\frac{iE}{\sqrt{ k}}} }{\sinh \frac{\pi E}{\sqrt{k}}} + 
\frac{2}{k} \left[ 4\pi \delta (E/\sqrt{k}) -  
\frac{\pi (\pi {\lambda})^{-\frac{iE}{\sqrt{ k}}}}{\sinh \frac{\pi E}{\sqrt{k}}}\right] 
J^3_{-1} \bar J^3_{-1} + \cdots \right) 
|j \, \nicefrac{E}{2}\,  \nicefrac{E}{2}\,  w \rrangle
\\
 \otimes 
| B\rangle_{\mathcal{M}} \otimes 
(1-\tilde b_{-1} c_{-1}-  b_{-1} \tilde{c}_{-1} + \cdots ) (c_0 + \bar c_0 ) 
c_1 \bar c_1 |0 \rangle_{gh} \, ,
\end{multline}
with contributions both from the continuous and discrete representations of 
$\mathfrak{sl} (2,\mathbb{R})$. Isolating the component proportional to 
$(\tilde{c}_{-1} J^3_{-1} + c_{-1} \bar J^3_{-1}) $ 
in~(\ref{brstbdy}) we find the conserved energy associated with the rolling tachyon 
in AdS$_3$, expressed in momentum space as follows:
\begin{equation}
T_{00} (j,w) \propto \frac{1}{g_s}\  
\nu_{k-2}^{1/2-j}  \,  
\frac{\Gamma \left( j+\frac{kw}{2} \right) 
\Gamma \left(j-\frac{kw}{2}\right)}{
\Gamma (2j-1) \Gamma (1+ \frac{2j-1}{k-2})} \, .
\label{energyprofile}
\end{equation}
Because the brane is point-like the stress-energy tensor has no components longitudinal 
to the brane worldvolume. We did not calculate the components transverse to the brane, that 
vanish in flat space-time. 

One can find the tachyon energy profile in space-time by inverse Fourier transform 
in the sector $w=0$. It is identical (up to a normalization constant) to the wave-function
corresponding to the D-particle in AdS$_3$. In the semi-classical limit, 
we find $T_{00} (\rho,\phi) \propto g_s^{-1} \, \delta (\rho)$,   
so the tachyon energy density is sharply localized and constant in global time. 
For finite $k$ the energy profile is smeared along the radial direction $\rho$ on a scale 
of order $\sqrt{\nicefrac{\alpha'}{k}}$.

\boldmath
\subsection{Rolling tachyon in AdS$_3$ from an orbifold construction}
\unboldmath
\label{subsecorbicons}
The coefficients of the boundary state for the rolling tachyon in AdS$_3$ --~or, in other words, 
the one-point function~-- can be obtained alternatively using an orbifold construction. 
This approach will prove later to be useful to analyze the open string sector of the theory. 
We start with a slightly unusual T-dual  representation of string theory on 
AdS$_3$ as the orbifold
\begin{equation}
\text{AdS}_3 \sim \frac{\slc^{\textsc{v}}_\infty \times \mathbb{R}^{0,1}}{\zi} \, .
\end{equation}
The first factor in this decomposition 
is the vector coset $\slc$ of the {\it universal cover} of $\slr$, i.e. the 
universal cover of the "trumpet" geometry.\footnote{It is T-dual  
to a $\zi$ orbifold of the "cigar"~\cite{Israel:2003ry}, i.e. of the axial coset. Another 
representation of this \textsc{cft} with a singular target space is 
the $\cn = 2$ Liouville theory with a momentum condensate, at infinite 
radius~\cite{Hori:2001ax}.} 
The latter has a metric 
\mbox{$\di s^2 = \di \rho^2 + \text{cotanh}^2 (\nicefrac{\rho}{\sqrt{2k}})\di x^2 $}
 with $x$ non-compact. 
In the coordinates \mbox{$\di s^2 = 2k\, \di z \di \bar z /(z \bar z -1)$}, one sees that it is  
conformal to the infinite cover of the exterior of the unit disc. 
The action of the $\zi$ translation orbifold on $\slc^{\textsc{v}}_\infty \times \mathbb{R}^{0,1}$,  
giving the AdS$_3$ \textsc{cft} is
\begin{equation}
\mathfrak{T}~: \quad (x,t) \longrightarrow 
(x+2\pi\sqrt{\nicefrac{2}{k}},t+2\pi\sqrt{\nicefrac{2}{k}}).
\label{acttrans}
\end{equation} 
Since it has no fixed points 
one can get the D-branes on AdS$_3$ from the branes of the 
\textsc{cft}  $\slc^{\textsc{v}}_\infty \times \mathbb{R}^{0,1}$ 
by summing over the images under the orbifold action. 

We start on the covering space of the orbifold with the tensor product 
of a D0-brane in the trumpet and a Neumann brane along the time 
direction with the rolling tachyon  
boundary deformation~(\ref{defsuper}). The former is an A-type brane of the 
vector coset $\slc$ which is 
in some sense localized on the boundary of the disc, in the strong coupling region,
in close analogy with the minimal length D1-branes of SU(2)/U(1) stretched between adjacent
"special points" on the boundary of the
disc~\cite{Maldacena:2001ky}. It carries a label $r_0$ giving its quantized 
position, with  $r_0 \in \zi$ for the universal cover of the manifold,\footnote{For the single cover of the 
trumpet $r_0 \in \zi_k$. It reflects 
the momentum non-conservation in the trumpet, T-dual to the winding non-conservation in the cigar.} 
with a coupling to the $x$-momentum of the coset (that we parameterize as $p_x = \mu \sqrt{2/k} $) 
in the one-point function of the form $\exp (4i\pi \mu r_0/k)$. 

We now construct the boundary state in the orbifold theory by summing over the images. The action 
of the translation generator $\mathfrak{T}$ on the labels of the brane is:
\begin{equation}
\mathfrak{T}~: \quad r_0 \longrightarrow r_0 +1 \quad , \qquad 
\tilde \lambda \longrightarrow \tilde \lambda e^{\pi \sqrt{\frac{2}{k}}} \, .
\label{translact}
\end{equation}
The transformation of the boundary cosmological constant $\tilde \lambda$ follows 
from the action of the translation~(\ref{acttrans}) 
on the boundary action~(\ref{defsuper}). We get 
the following boundary state (for the \textsc{ns-ns} sector):
\begin{multline}
|B\rrangle = \frac{1}{k\pi} \sum_{h \in \zi} 
\int \di j \, \di \mu \  \nu_k^{\nicefrac{1}{2}-j}  
 \  \frac{\Gamma (j+\mu) 
\Gamma (j- \mu)}{\Gamma (2j-1) \Gamma (1+ \frac{2j-1}{k})}\ e^{\frac{4i\pi \mu h}{k}}\ 
\times
\\ \times \ \frac{1}{\sqrt{2k\pi}}
\int \frac{\di E}{2\pi}
\left(\frac{\pi \tilde{\lambda}}{2}  e^{\pi \sqrt{\frac{2}{k}} h}
\right)^{-i\sqrt{\frac{2}{k}}E}\
\frac{\pi}{ \sinh 
\frac{\pi E}{\sqrt{2k}}}\,  |j,\mu,\mu\rrangle \otimes |E \rrangle \\
\end{multline}
where $|j,\mu,\mu\rrangle$ and $|E\rrangle$ are respectively the 
Ishibashi states for the $\slc$ vector super-coset with type A 
boundary conditions and for the time-like direction (with the same normalization 
$\Delta = -\nicefrac{E^2}{4k}$ as in AdS$_3$). 
Now performing the sum over $h$ (i.e. over the images under the orbifold 
action) one finds a boundary state whose coefficients are identical, up to an
overall normalization, to the one-point 
function~(\ref{adstachonept}) for the rolling tachyon in AdS$_3$.

\section{Closed and open string emission by the rolling tachyon}
\label{secemission}
The physics of the rolling tachyon is, from the closed strings point of view,
the decay of an unstable, very massive particle, producing radiation of closed 
strings composed of all the string modes coupling to the brane. 
As recalled in the introduction 
the average number of closed strings emitted by the 
D0-brane decay in flat space-time, computed in the tree-level approximation,
is divergent~\cite{Lambert:2003zr}. This is the detailed outcome of
the competition between the exponential suppression of the emission
probability at high energy, and the exponential growth of the density of 
string modes. Physically, because 
the initial energy is finite and equal to the mass of the D-particle of 
order $\nicefrac{1}{\sqrt{\alpha'} g_s}$,  
one must impose a cutoff at the corresponding energy, which is exactly the value 
for which non-perturbative effects kick in. Then one concludes that 
the preferred decay channels are non-relativistic 
very massive excited closed strings.  The decay of higher dimensional 
branes produces a ``tachyon dust'' of pressure-less fluid.

In AdS$_3$ we may expect a different picture, as the density of states 
at high energy behaves like a field theory (because  
string theory in AdS$_3$ is dual to a two-dimensional 
conformal field theory).   We  show below that 
"long strings" play a prominent role in the closed strings description of the decay. 
We identify non-perturbative effects that do regularize ultraviolet divergences 
and give finite physical quantities resulting from the decay. This however 
does not challenge Sen's conjecture~\cite{Sen:2003iv} since all the D0-brane energy 
is converted into closed strings radiation, as will be explained in this section.

\subsection{Annulus amplitude for the decaying D-particle}
The mean number of emitted closed strings can be obtained from  
the imaginary part of the annulus diagram, 
using an optical-like theorem and open/closed channel duality. We follow closely 
the computation of~\cite{Karczmarek:2003xm}, and the recent 
analysis of~\cite{Nakayama:2006qm} that clarified 
the issue of the analytic continuation. It gives also an analogue of the Cardy 
consistency condition~\cite{Cardy:1989ir,Teschner:2000md} for the boundary state.

The annulus amplitude in the closed string channel, 
computed from the one-point function,  
gets contributions both from the continuous and from the 
discrete representations. 
For simplicity of the notations in the following manipulations we keep track only of the
continuous representations in the \textsc{ns} sector; one can  
check that the contributions of the discrete representations 
and the \textsc{r} sector of the full superstring background are 
consistent with the open string annulus amplitude~(\ref{tachpart}) 
that we get at the end of the computation. From the expression of the 
one-point function, see eqn.~(\ref{adstachonept}),
we obtain the closed string channel integrand of the  
annulus amplitude, in the \textsc{ns} sector of continuous representations 
as (with $\tilde q = \exp{ 2i\pi \tilde{\tau}}$):
\begin{multline}
Z (\tilde \tau )  =  \frac{i\pi}{2} \sqrt{\frac{2}{k}}  \sum_{w\in \zi} \int_{-\infty}^{\infty} \frac{\di
  E}{2\pi}  \
\int_0^\infty \di P \ \frac{\sinh 2\pi P \sinh \frac{2\pi P}{k}}{\left[
\cosh 2\pi P+\cos \pi(E-kw)\right]\sinh^2  
\frac{\pi E}{\sqrt{2k}}} \ \times  \\ 
\times \ ch_c \oao{0}{0} \left(P,\frac{E-kw}{2}  ;
  \tilde{\tau} \right) \ 
\frac{\tilde{q}^{-\frac{E^2}{4 k}}}{\eta (\tilde \tau )} \left(
  \frac{\vartheta \oao{0}{0} (\tilde \tau)}{\eta (\tilde \tau)} \right)^{1/2} \, . 
\label{closechanrolads}
\end{multline}
To obtain a well-defined 
expression we should consider, as explained in detail in~\cite{Nakayama:2006qm}, 
the Lorentzian cylinder with the appropriate $\varepsilon$-prescriptions, i.e. take the modular
parameter $\tilde \tau = 1/t+ i \varepsilon$, $t \in \mathbb{R}_+$ and
regularize the integration over the energy using the prescription
\mbox{ $E \to E_{-\hat{\varepsilon}} \equiv (1-i\hat{\varepsilon})E$}.

To be able to modular transform to the open string channel we need to follow 
some algebraic manipulations. We will extract at the end the imaginary part of the 
annulus amplitude that 
can be interpreted as the closed string emission. The first step of the
computation is to disentangle the $\slc$ part and the $U(1)$
part of the closed strings channel amplitude. We rewrite~(\ref{closechanrolads}) as follows
{\allowdisplaybreaks 
\begin{subequations}
\begin{align}
Z (\tilde \tau )  = & \ \frac{i}{4} \left(\frac{2}{k}\right)^{3/2}  \sum_{w} \int \di \mu\,  \di E
\ \int_0^\infty \di P \ \frac{\sinh 2\pi P \sinh \frac{2\pi P}{k}}{\left[
\cosh 2\pi P+\cos 2\pi \mu \right]\sinh^2  \frac{\pi E}{\sqrt{2k}}}
 \  \times 
\nonumber \\ 
& \times \ ch_c \oao{0}{0} \left(P,\mu  ; \tilde \tau  \right) \ 
\frac{\tilde{q}^{-\frac{E_{-\hat{\varepsilon}}^2}{4 k}}}{\eta (\tilde \tau )} \ 
\delta \left( \frac{2\mu}{k} + w - \frac{E}{k} \right)
\left(\frac{\vartheta \oao{0}{0}(\tilde \tau)}{\eta (\tilde \tau )}\right)^{1/2} \\
& \nonumber \\
= & \  \frac{i}{4} \left(\frac{2}{k}\right)^{3/2}   \int \di \mu  \sum_{r \in \zi}
\int_0^\infty \di P \ \frac{\sinh 2\pi P \sinh \frac{2\pi P}{k}}{
\cosh 2\pi P+\cos 2\pi \mu} \ e^{-\frac{4i\pi r\mu}{k}} ch_c \oao{0}{0} \left(P,\mu  ;
  \tilde \tau  \right)
\ \times \nonumber \\ 
& \times \ \int \frac{\di E}{  
\sinh^{2} \left( \nicefrac{\pi E}{\sqrt{2k}} \right)}   \ 
e^{+\frac{2i\pi r E}{k}}\ 
\frac{\tilde{q}^{-\frac{E_{-\hat{\varepsilon}}^2}{4 k}}}{\eta
  (\tilde \tau )}  \left( \frac{\vartheta \oao{0}{0}(\tilde \tau)}{\eta (\tilde \tau) } \right)^{1/2}  . 
\label{intcylampl} 
\end{align}
\end{subequations}
We recognize in the first line  of~(\ref{intcylampl}) the continuous part of 
the modular transformation from the identity representation characters of 
$\slc$~\cite{Eguchi:2003ik,Ahn:2003tt,Israel:2004jt} (with $\tau = -\nicefrac{1}{\tilde \tau} 
$):
\begin{equation}
ch_\mathbb{I}\oao{0}{0} (r;\tau ) = \frac{4}{k}  
\left\{
\int_{0}^{\infty} \di P \! \int \di \mu \, e^{-4i\pi \frac{\mu r}{k}}
\frac{ \sinh 2\pi P  \sinh \frac{2\pi P}{k}}{\cosh 2\pi P + \cos 2\pi \mu}
 ch_{c} \oao{0}{0} (P,\mu; \tilde \tau) +  \text{discrete} \right\}
\label{trivialtrans}
\end{equation}}
where the second term contains an integral over characters in the discrete representations
of $\slc$ with spin in the range $\nicefrac{1}{2}<j<\nicefrac{k+1}{2}$. 
One can check that this discrete part of the modular transform matches 
the contribution of the discrete representations 
to the closed string channel annulus amplitude that we get from residues at 
the poles of the one-point function~(\ref{adstachonept}). 

Then we have to deal with the U(1) part of~(\ref{intcylampl}), i.e. 
the integral over space-time energy. In terms of the open string channel  
modulus $\tau = t - i \varepsilon$ (regularized by the $\varepsilon$ prescription), 
we get the  modular 
transformation (with $q = \exp 2i\pi \tau$): 
\begin{equation}
\int \frac{\di E}{\sinh^{2} \left( \nicefrac{\pi E}{\sqrt{ 2k}} \right)}\ 
e^{\frac{2i\pi r E}{k}}\ 
\frac{\tilde{q}^{-\frac{E_{-\hat{\varepsilon}}^2}{4 k}}}{\eta (\tilde \tau )} 
=   \int \di \upsilon  \ 
\frac{q^{-\frac{\upsilon_{\hat{\varepsilon}}^2}{2}}}{\eta ( \tau )} 
\int \di E \ \frac{\cos 
\frac{2\pi E}{k}  
\left( \sqrt{\frac{k}{2}}\, \upsilon + r \right)}{
\sinh^{2}  \left( \nicefrac{\pi E}{\sqrt{ 2k}}\right)} \, , 
\end{equation} 
with as before $\upsilon_{\hat{\varepsilon}}\equiv (1+i\hat \varepsilon) \upsilon$. 
Finally, adding the contribution from $\slc$ discussed above we get the 
open string channel amplitude as:
\begin{equation}
Z (\tau ) = \frac{i}{8} \sqrt{\frac{2}{k}} \sum_{r \in \zi} ch_\mathbb{I} (r; \tau ) \ 
\ \int \di \upsilon \ \frac{q^{-\frac{\upsilon_{\hat{\varepsilon}}^2}{2}}}{\eta (\tau)} \times 
\left[ \int \di E \ \frac{\cos \frac{2\pi E}{k}  
\left( \sqrt{\frac{k}{2}}\, \upsilon +   r \right)}{
\sinh^{2}  \left( \nicefrac{\pi E}{\sqrt{2k}}\right)} \right] \, .
\end{equation}
As expected, the density of states given by the bracketed expression 
is divergent (it has a double pole at $E=0$); this infrared divergence is 
due to the infinite ``volume'' along the time direction as 
in ordinary Liouville theory. We can subtract 
this double pole, that will not contribute to the imaginary 
part of the amplitude, and then introduce the special 
function $S_\beta^{(0)} (x)$~\cite{Fateev:2000ik}:
\begin{equation}
\ln S_\beta^{(0)} (x) = \frac{1}{2} \int_0^\infty \frac{\di y}{y} \ 
\left[ \frac{\sinh (Q_\beta -2x)y }{\sinh \beta y \ \sinh \beta^{-1} y} 
+ \frac{2x-Q_\beta}{y} \right] \, ,
\label{Sfct}
\end{equation}
with $Q_\beta = \beta + \beta^{-1}$. In terms of this function 
we obtain the open string channel annulus amplitude for the decaying 
D-particle in AdS$_3$ as follows:
\begin{equation}
Z (\tau ) =   
\sum_{r \in \zi} ch_\mathbb{I}
\oao{0}{0} (r; \tau ) 
 \int  \di \upsilon \ \frac{1}{2\pi} \frac{\di}{\di \upsilon} 
 \ln S_1^{(0)} \left(1- i\upsilon- 
i\sqrt{\frac{2}{k}}\, r \right) 
\frac{q^{-\frac{\upsilon_{\hat{\varepsilon}}^2}{2}}}{\eta (\tau)}  \left( \frac{\vartheta \oao{0}{0} (\tau
    )}{\eta (\tau)} \right)^{\nicefrac{1}{2}} \, . 
\label{spectraldensads}
\end{equation}
Let us now consider the physical setup of interest, 
namely type \textsc{iib} strings on 
\mbox{AdS$_3 \times$ S$^3 \times$ T$^4$} in the presence of the D-particle. 
Let us remind that we have chosen to take a D0-brane in the S$^3$ part of the
background, i.e. the symmetric brane of SU(2) with $\hat \jmath = 0$ 
(which implies that only the $\tilde \jmath =0$ representation 
is possible in the open string sector) and Dirichlet boundary conditions along all the cycles of the 
four-torus.  There are also extra contributions from the 
twisted Neveu-Schwarz (i.e. with a $(-)^F$ insertion) and 
Ramond sectors to the annulus 
amplitude in type \textsc{ii} superstrings. 
To write the density of states in the open string channel twisted 
$\textsc{ns}$ sector (coming from the closed string channel 
\textsc{r} sector) 
it is useful to define another special function, which  turns out to 
have no pole to subtract at the origin: 
\begin{equation}
\ln S_\beta^{(1)} (x) = \frac{1}{2} \int_0^\infty \frac{\di y}{y} \ 
\frac{\sinh (Q_\beta -2x)y }{\cosh \beta y \ \cosh \beta^{-1} y} \, .
\end{equation}
Once adding everything and integrating over the modular parameter $\tau$ we get 
the integrated annulus amplitude for the rolling tachyon 
in AdS$_3 \times$ S$^3 \times$ T$^4$:
\begin{multline}
\mathcal{A} = \frac{1}{2} \sum_{a,b \in \zi_2}(-)^{b}  \int \frac{\di t}{2t}
\sum_{r \in \zi} ch_\mathbb{I} \oao{b}{a} (r;\tau) \ \times  \\
\times \
\int \di \upsilon \  
  \frac{1}{2\pi} \frac{\di}{\di \upsilon} \ln S_1^{(a)} \left(1- i\upsilon - i\sqrt{\frac{2}{k}}(r+b/2) \right) \ 
q^{-\frac{\upsilon_{\hat{\varepsilon}}^2}{2}} 
\chi^{0} (\tau )   \sum_{\{ w^i\}} 
\frac{q^{\frac{(R_i w^i)^2}{2}}}{\eta^3 (\tau )} \ 
\frac{\vartheta \oao{b}{a}^3 (\tau )}{\eta^3 (\tau )} \, .
\label{tachpart}
\end{multline} 
From this result we will derive in the following subsection 
the production of closed strings as the imaginary 
part of the annulus amplitude.

As in ordinary Liouville theory the "energy density" 
$\p_\upsilon \ln S_1^{(a)} (1- i\upsilon - i r \sqrt{2/k}\, )$ is  related 
to the boundary two-point function~\cite{Teschner:2000md}. This provides
a consistency check for the boundary state, analogous to the Cardy condition. 
The annulus amplitude contains first a contribution from the 
``tachyon sector'', i.e.  open strings with odd fermion number 
whose lowest state is the open string tachyon. The associated 
spectral density is written in terms of 
\begin{multline}
\ln S_\beta^\textsc{t} (x) =  
\frac{1}{2}\left( \ln S_\beta^{(0)} (x) + \ln S_\beta^{(1)} (x)  \right)\\
=  \int_0^\infty \frac{\di y}{y} \left[ 
\frac{\sinh (\frac{Q_\beta}{2}-x)y \ \cosh \frac{Q_\beta y}{2}}{\sinh \beta y \sinh \beta^{-1} y} + \frac{x-\frac{Q_\beta}{2}}{y} \right] 
 =  \ln  S_\beta^{(0)} \left(\frac{x}{2}\right) S_\beta^{(0)}
\left(\frac{x+Q_\beta }{2} \right) \, . 
\label{denstach}
\end{multline}
This expression is closely related to the boundary two-point function 
of super-Liouville theory given in~\cite{Fukuda:2002bv}. We will discuss 
this issue in more detail in subsection~\ref{openpair}.  Similarly for the 
``vector sector'', i.e. open string states with even fermion number, 
the spectral density is written using
\begin{equation}
 \ln S_\beta^\textsc{v} (x) = \frac{1}{2}\left( \ln S_\beta^{(0)} (x) - \ln S_\beta^{(1)} (x)  \right)=
 \ln  S_\beta^{(0)} \left(\frac{x+\beta}{2}\right) S_\beta^{(0)}
\left(\frac{x+\beta^{-1} }{2} \right) \, .
\label{densvec}
\end{equation} 
For the Ramond sector a similar decomposition according to the fermion number can be carried 
out; however it is not useful in order to analyze the annulus amplitude since the contribution 
from the twisted Ramond sector is identically zero.

\subsection{Analysis of the closed string emission}
As in~\cite{Karczmarek:2003xm, Nakayama:2006qm} we will now extract the
imaginary part of this annulus amplitude, that appears once we Wick-rotate  
the expression~(\ref{tachpart}) to Euclidean signature 
$\upsilon \to i \upsilon$.\footnote{Together 
with a Wick rotation of the Schwinger parameter $t$. } It gives the mean number of 
produced on-shell closed string states as we will check 
later by interpreting this quantity in the closed string channel.  

The imaginary part of the annulus diagram  comes from the simple poles of 
$S_\beta^{(0)} (x)$ in the densities~(\ref{denstach},\ref{densvec}) (and also
in  the \textsc{r} sector),  located at $x= -n\beta-m\beta^{-1}$, 
for $m,n \in \zi_{\geqslant 0}$, and the simple zeroes 
for  $m,n \in \zi_{< 0}$. For $\beta=1$ both give double poles 
of the energy density. Compared to the flat space-time computation, with poles only 
on the imaginary axis, the partition function contains new poles all over the 
$\upsilon$-plane 
in the sectors $r \neq 0$, much like in the annulus for the 
accelerating D-brane in NS5-backgrounds  studied in~\cite{Nakayama:2006qm}. 
The residues of the poles situated in the upper-right and lower-left quadrants
of the $\upsilon$-plane, including the imaginary axis, will then contribute 
to the imaginary part of the annulus amplitude. Note finally that these 
poles don't correspond to on-shell physical open string states; the associated 
open string wave-functions decrease exponentially with time towards past  infinity. 
They correspond rather to on-shell closed string states as will be shown below using 
channel duality.

Adding the contributions from the 
different sectors and using the symmetry~(\ref{reflsym}) defined 
in appendix~\ref{appeads}, we obtain from~(\ref{tachpart}) 
the imaginary part of the annulus amplitude, written 
in the open string channel, as the sum over the residues:
\begin{multline}
\text{Im} \, \mathcal{A}  =  \frac{1}{2} \int_0^\infty \frac{\di t}{2t}
\sum_{a,b \in \zi_2}(-)^b 
\chi^{0} (it )\ \frac{\vartheta \oao{b}{a}^3 (it )}{\eta^3 (it )} \sum_{\{ w^i\}} 
\frac{e^{-\pi t (R_i w^i)^2}}{\eta^3 (it )} 
 \sum_{r \in \zi} ch_\mathbb{I} \oao{b}{a} (r;it)
\ \times  \\ \times \ 
\sum_{n=1}^{\infty}
 (-)^{a(n+1)}\, n \,  e^{-\pi t \left( n-
i\sqrt{\frac{2}{k}}(r+b/2) \right)^2} \, .
\label{imannul}
\end{multline}
The sign $(-)^{a(n+1)}$ in the last line comes from the analysis of the 
poles in the open string twisted $\textsc{ns}$ sector. 
Note that the sum over $r$ is convergent because 
we get a factor of $e^{-2\pi t \{(r+b/2)^2/k+|r|\}}$ from the identity character 
$ch_\mathbb{I} (r;it)$ of $\slc$. As we show below, 
this expression gives the distribution of closed string emission, 
as it is expected  using an optical-like theorem and 
open/closed string duality of the annulus diagram~\cite{Karczmarek:2003xm}. 
To obtain a proper closed strings interpretation of this quantity we need 
to modular transform back to the closed string channel; the computation is 
given in appendix~\ref{annulustransfo}. 

\subsubsection{Long strings production}
We show that the mean number of produced closed strings is equal to the imaginary 
part of the annulus diagram given by eqn.~(\ref{imannul}), looking
first at the emission of long closed strings. The initial condition 
for closed strings at past infinity  set by the rolling tachyon worldsheet  \textsc{cft} 
corresponds to the non-\textsc{bps} D0-brane boundary state, in the absence of closed string radiation.  
We may expect that the production of very massive long strings is highly suppressed, 
because the density of states at high energies is significantly lower than in flat 
space-time (because the energy scales like $E\sim N/w$ with the oscillator number $N$, 
compared to $E\sim \sqrt{N}$ in flat space-time). 
We turn now to the exact computation that partially confirms these expectations. 

After a  modular transformation of the imaginary part of the annulus amplitude~(\ref{imannul}) 
to the closed string channel, one get first a contribution from the continuous $\slr$ representations:
\begin{multline}
\text{Im} \mathcal{A}_c  = 
\frac{1}{4k \prod_i R_i} \sum_{|\phi\rangle \in \mathcal{H}_c }\sum_{2\tilde \jmath=1}^{k-2} 
\sin \left( \frac{\pi (2\tilde \jmath +1)}{k} \right) \sum_{\{ n_i \} \in \zi}
 \int \di P  \sum_{a,b} \sum_{N}(-)^{bF}  D(N) \ \times \\ \times    
\sum_{w \neq 0} \frac{1}{|w|}\ \left[
\frac{ \sinh 2\pi P  \sinh \frac{2\pi P}{k}}{\cosh 2\pi P + 
\cos \pi (kw -E_a)}  \right]
 \frac{(-)^a}{\sinh^2 \left(\frac{\pi E_a}{\sqrt{2k}}+ \frac{i\pi a}{2}\right)} 
\label{prodlongexpl}
\end{multline}
where the sum runs over the physical states $|\phi\rangle$ in sub-Hilbert space 
$\mathcal{H}_c$ of closed strings made with $\slr$ continuous representations,  
that couple to the brane. 
In this equation $\tilde \jmath$ is the SU(2) spin, $n_i$ are the toroidal momenta, 
$D(N)$ is the density of states at oscillator number $N$  and $F$ 
the worldsheet fermion number for a given string state. All these 
labels have to be understood as functions of the    
state $|\phi\rangle$.\footnote{We should 
in principle take care more carefully of the different 
contributions to the density of states, in particular of the null
states. However we will eventually be interested in the high-energy tail 
of the distribution for which these details are irrelevant.} 
The term between square brackets is the contribution from $\slc$ to the amplitude, while 
the last factor corresponds to the amplitude for the time-like boson part. 
The on-shell space-time energy $E_a$ 
for a long string with $w$ units of spectral flow and radial momentum $P$, 
in the \textsc{ns} ($a=0$) or \textsc{r} ($a=1$) sector, reads~\cite{Maldacena:2000hw}:
\begin{equation}
E_a = \frac{kw}{2} +\frac{2}{w} \left[  \frac{P^2+\nicefrac{1}{4}}{k}+
\frac{\tilde \jmath(\tilde \jmath+1)}{k}  + \frac{\sum_i(n_i/R_i)^2}{2}+ N+\frac{a-1}{2}\right]\, ,
\end{equation} 
Only left-right symmetric states couple to the brane. There are also 
no on-shell continuous representations for $w=0$. The details of the computation are given in appendix~\ref{annulustransfo}. 

This quantity is related to the closed string emission by the brane decay. 
In terms of the coefficient $\mathcal{V}^{w} ( E,P,\cdots)$ of the
one-point function for a  closed string vertex operator  in the 
rolling tachyon background, eqn.~(\ref{adstachonept}), it can be expressed as 
{\allowdisplaybreaks
\begin{align}
\text{Im} \mathcal{A}_c &= 
 \frac{1}{2\pi} \sum_{|\phi\rangle} \int \frac{\di P}{\pi} \, \frac{1}{2}\sum_{a,b} 
\sum_{w \neq 0 ,N,\ldots} (-)^{bF} D(N) \frac{1}{|w|} \left| \langle 
\mathcal{V}^{w} \left( E_a \left[P,w,N,\ldots \right],P,\cdots \right) \rangle 
\vphantom{\frac{1^1}{2_2}} \right|^2  \nonumber \\
& = \sum_{|\phi\rangle}
\int \frac{\di E}{2\pi} \int_0^\infty \frac{\di P}{\pi}\, 
\frac{1}{2}\sum_{a,b}\sum_{w,N,\ldots}  (-)^{bF} D(N) \ \times \nonumber \\ & \qquad \qquad
\qquad \times \ \delta \left(-wE +\frac{kw^2}{2}+ \frac{2(P^2+\nicefrac{1}{4})}{k}+2N+\cdots \right)
\left| \langle \mathcal{V}^{w} (E,P,\cdots) \rangle \right|^2 .
\label{longmean}
\end{align}}
Following~\cite{Lambert:2003zr}, this amplitude can be interpreted as 
$\bar{\mathcal{N}}_c$, the mean number of long strings produced by 
the D0-brane decay, traced over all the physical spectrum of closed 
strings in the continuous representations.
We work in the gauge with no oscillators for the time-like boson, 
which is consistent as long as spacetime energy is not zero~\cite{Hwang:1991an}.\footnote{
Also the amplitude for a properly normalized descendant
of the $\cn =2$ superconformal algebra in  the $\slc$ super-coset 
is the same, up to a phase, as the amplitude of the primary state. It 
can be shown using the (type B) boundary conditions on the $\cn =2$ \textsc{sca} 
generators.} In the present case there are no physical long strings with zero 
energy. The invariant measure that appears in~(\ref{longmean}), with the on-shell constraint 
$\delta (L_0 +\bar L_0 +a-1)$, is a natural generalization
of the point-particle measure to string theory. 

The absence of a delta-function representing the conservation of the {\it total} energy stored in the brane, 
which is justified for $g_s = 0$, is a consequence of the fact that eqn.~(\ref{longmean}) is a 
tree-level amplitude (indeed one applies an optical-like theorem to the one-loop annulus amplitude).

\paragraph{High energy behavior}
We would like now to check whether, like in flat space-time, 
there is an ultraviolet 
divergence associated with the production of very massive closed strings. 
The leading term in the density of states, for the 
left-right symmetric string states that couple to the D-brane, is given 
by~\cite{Cardy:1986ie,Kutasov:1990sv}:
\begin{equation}
\ln D(N) = 2\pi \sqrt{\frac{c_\text{eff} N}{6}} + \mathcal{O}(\ln N )\, ,
\label{cardy}
\end{equation}
where the effective central charge is $c_\text{eff} = 12 -
\nicefrac{6}{k}$. To be more precise one should count the transverse physical degrees 
of freedom in space-time using the decomposition $\slr \sim \slc \times \mathbb{R}^{0,1}$. 
Then the effective central charge differs from the central charge by an amount 
proportional to the minimal scaling dimension in the spectrum of the coset. Alternatively 
the asymptotic degeneracy of states can be computed directly from the 
covariant AdS$_3$ partition function for the continuous
representations, since their characters are identical to those of a free field theory, 
leading to the same result.\footnote{Another way to reach  
this conclusion is to consider the effective linear dilaton \textsc{cft} living 
on the long strings~\cite{Giveon:2005mi}.}  
As in~\cite{Karczmarek:2003xm}, the difference between the 
central charge and the effective central charge decreases the density of states at high
energy compared to flat space-time,  but in AdS$_3$ the main result is that 
there is no Hagedorn density of states in a given sector of spectral flow $w$.

Indeed,  the spacetime energy of long strings, 
in the spectral flow sector $w$, grows with the oscillator number as 
$E \sim \nicefrac{2N}{w}$. Therefore the suppression of the production 
of very energetic strings wins over the growth of the density of states 
which scales only like $\sqrt{N}$ as we saw above. It shows that, for a given 
sector of spectral flow, the production of very massive closed strings 
is exponentially suppressed.

Let us first look at the distributions of the closed string radiation for the 
radial momentum $P$ and the oscillator number $N$, in a given sector of 
spectral flow $w$. In~(\ref{prodlongexpl}), the leading parts of the $P$-dependence 
from the $\slc$ contribution cancel for large $P$, leaving an exponential contribution 
of the form $\exp ( \nicefrac{2\pi P}{k})$.  Omitting 
for the moment the summation over the zero modes of S$^3 \times $T$^4$, 
the amplitude~(\ref{prodlongexpl}) for large $N$ and $P$ behaves like:
\begin{multline}
\bar{\mathcal{N}}_c \sim \frac{1}{k} \sum_{w =1}^{\infty} \frac{1}{\omega}\, e^{-\pi \sqrt{\frac{k}{2}} w} 
\left\{ \int_0^\infty \di P \ e^{-\frac{2\pi}{k}\left[ \sqrt{\frac{2}{k}}
\frac{P^2}{w}  
-P \right]}\right\} 
\sum_{N} N^\gamma e^{-2\pi \left( \sqrt{\frac{2}{k}}\frac{N}{w} 
-\sqrt{(2-\nicefrac{1}{k})N}\right)} \\
\label{longbehav}
\end{multline}
Notice again the term in the second exponential proportional to  $N/w$ which 
differs from that similar term in flat space-time (or in the presence of a linear dilaton) 
where only $\sqrt{N}$ would have appeared. The exponent $\gamma$ comes from the logarithmic corrections 
to the asymptotic density of states~(\ref{cardy})~\cite{Abel:1999rq}. 
The distributions in radial momentum $P$ and oscillator number $N$ 
for a given sector of spectral flow $w$ are centered at 
\begin{equation}
\bar P_w  = \sqrt{\frac{k}{2}}\frac{|w|}{2} \quad , \qquad \bar N_w = 
\frac{w^2 (2k-1)}{8} \, ,
\label{meanval}
\end{equation}
see figure~\ref{curvesdist}. 
We see that the emitted long strings with large winding number 
are very excited; they have an energy of order $E\sim kw/2+2\bar
P_w^2/(kw)+2\bar N_w /w=kw$. The radius of a long string with radial 
momentum $P$ and winding number $w$ growths linearly in 
global AdS$_3$ time as~\cite{Maldacena:2000hw}
\begin{equation}
\rho = \frac{2 P}{k} \left| \frac{t}{w}\right|\, .
\end{equation} 
Plugging in this formula the mean value of $P$, eqn.~(\ref{meanval}), 
we observe that the mean speed of growth of 
the long strings with large $w$ emitted by the D0-brane decay is independent 
of their winding number, with a standard deviation of order 
$k^{-1/4}w^{-1/2}$.

\FIGURE{
\includegraphics[width=120mm]{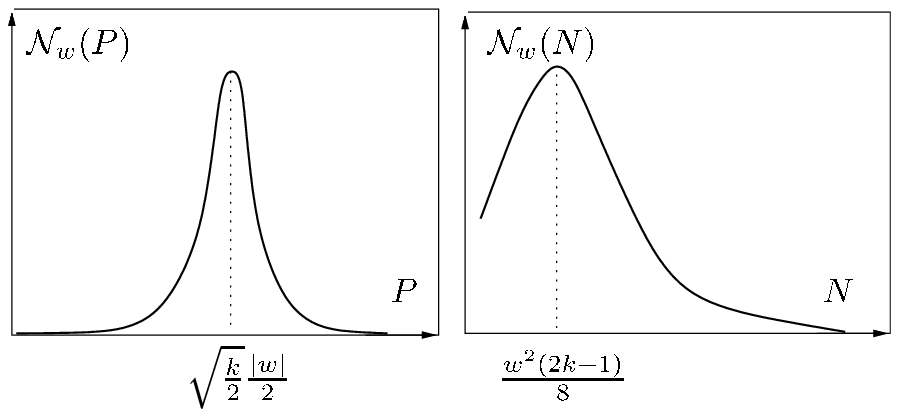}
\caption{\it Distribution of the long strings radiation for the radial
momentum $P$ and the oscillator number $N$ in a given sector of spectral flow $w$.}
\label{curvesdist}
}

To check whether the sum over the different sectors of spectral 
flow $w$ gives a finite number of produced long strings, we first 
integrate~(\ref{longbehav}) over $P$; indeed 
the term   $\exp ( \nicefrac{2\pi P}{k})$ can enhance the production of 
very  massive long strings with large radial momentum. We get:
\begin{equation}
\bar{\mathcal{N}}_c \sim 
\sum_{w>0} \frac{1}{\sqrt{w}}\ e^{-\pi \sqrt{\frac{k}{2}}\left(1-\frac{1}{2k}\right)w} 
\sum_{N} N^\gamma  
e^{-2\pi \left( \sqrt{\frac{2}{k}}\frac{N}{w}
    -\sqrt{(2-\nicefrac{1}{k})N}\right)}. 
\label{behavintp}
\end{equation}
Then in the 
large $w$ limit one can replace  the sum over the oscillator number $N$ by 
an integral. We find that the exponential terms in $w$ cancel 
from the amplitude~(\ref{behavintp}).  Therefore, the distribution of the radiation as a 
function of  $w$ is governed by the power-law corrections to the asymptotic density 
of states, much as in flat space-time.

To evaluate the effect of the power-law corrections, we  
consider first the simplest case of a non-critical superstring 
$\slr |_{k=2} \times $S$^1 \times \mathbb{R}^4$, with the brane localized on 
$\mathbb{R}^4$:
\begin{equation}
\bar{\mathcal{N}}_c \sim \sum_{w>0} \frac{1}{\sqrt{w}}\, e^{-\pi
   \left(1-\frac{1}{4}\right)w} 
\sum_{N} N^\gamma  
e^{-2\pi \left( \frac{N}{w}
    -\sqrt{3N/2}\right)}
\left[ \int \di p\  e^{-\pi\frac{p^2}{w}} \right]^4
\label{longbehavnc}
\end{equation}
First, integrating over the $\mathbb{R}^4$ momenta 
one gets an extra factor of $w^2$ in the sum~(\ref{longbehavnc}). 
Then, we have to count the contributions to the 
density of states of five free bosons and eight free fermions (two of them 
being the S$^1$ at the fermionic radius).  Using the results 
of~\cite{Carlip:2000nv} we find a "universal" correction 
of order $N^{-3/4}$, and an extra contribution of order 
$N^{-5/4}$ from the five bosonic oscillators (in the light-cone gauge).  
Thus in this case the exponent of the power-law correction is $\gamma=-2$ and  the sum~(\ref{behavintp})  
behaves like $\bar{\mathcal{N}}_c \sim \sum_{w>0} w^{-1}$. One can check that if one compactifies 
$\mathbb{R}^4$, for example on a square torus at the fermionic radius for which the computation is simple, 
this result holds.

We find that 
the total number of emitted long strings from the decaying D0-brane 
is log-divergent. Their mean energy is linearly divergent in $w$. It shows that 
the ultraviolet divergence of closed string emission is not removed in AdS$_3$, despite the much 
softer high-energy behavior.\footnote{Technically the factor $\sinh 2\pi P/k$ in~(\ref{prodlongexpl}) 
responsible for this result comes from the (absolute square of the) worldsheet non-perturbative 
corrections to the $\slc$ one-point function, i.e. the term $\Gamma(1+\frac{2j-1}{k})$ 
in~(\ref{adstachonept}).} Closed string perturbation theory breaks down due 
to a large backreaction from the long strings. The same conclusions can be reached for 
the other non-critical string backgrounds   AdS$_3 \times$ S$^1 \times$ 
T$^2$  and AdS$_3 \times$ S$^1$. In appendix~\ref{bosprod}
we discuss in detail the bosonic case where the states counting can be done  
explicitly. In the more generic cases like superstrings on \mbox{AdS$_3 \times$ S$^3 \times$ T$^4$} 
the computation of the power-law corrections to the density of states 
is more involved but will very likely leads to the same
conclusion. At least it is safe to say that, because of the power law
behavior, higher moments $\overline{E^n}$ will be divergent for $n$ large enough, 
signaling the breakdown of closed string perturbation theory.

\paragraph{Non-perturbative regularization} The tree-level computation of closed strings 
radiation predicts that an infinite amount of energy is emitted by the decay of the 
D0-brane. This divergence is regularized once non-perturbative effects kick in. 
The spectral flow is non-perturbatively bounded from above, because 
long strings with winding number $w \sim Q_1$ carry an \textsc{ns-ns} two-form charge of the same order as the 
total charge of the $Q_1$ background fundamental strings, for an AdS$_3 \times$ S$^3 \times$ T$^4$ superstring 
with coupling constant $g_6 = \sqrt{k/Q_1}$. This regularization gives 
$\bar{\mathcal{N}}_c \sim -\log g_6^2$ and the average radiated energy as 
$\bar E \sim Q_1 \sim \nicefrac{1}{g_6^2}$. At weak coupling is far larger than 
the mass of the brane which is of order $\nicefrac{1}{g_6}$. 
 
One can understand this discrepancy, at least qualitatively, as follows. 
The amplitude contains contributions from long strings with energies larger than 
(in units of $1/\sqrt{\alpha'}$) $\nicefrac{1}{g_6}$, the inverse 
six-dimensional string coupling. One may make the assumption that the contribution to 
the amplitude of these long strings, 
which lie outside the domain of validity of string perturbation theory,  
is at most subdominant (their energy is of order of, or larger than, the 
D0-brane mass). Therefore, since the typical energy of long 
strings with winding $w$ is $E = kw$, to exclude strings with 
$E > 1/g_6$ one needs to put a cutoff of order $w < g_6^{-1} \propto \sqrt{Q_1}$ on 
the spectral flow number $w$. While from the point of view of the number of produced long 
strings the effect is not dramatic, it gives an average radiated energy of order 
$\nicefrac{1}{g_6}$,  in agreement with energy conservation. 
The distribution of closed strings radiation is peaked near the cutoff scale.

One might have expected a qualitative 
difference for $k<1$,\footnote{
For the AdS$_3 \times$ S$^3 \times$ T$^4$ 
background that we take as an example we have always $k\geqslant 2$
but the regime $k<1$ can be obtained for certain non-critical
backgrounds~\cite{Giveon:2005mi}.} because string theory on AdS$_3$ undergoes a
phase  transition there and long strings become weakly coupled near the
boundary~\cite{Giveon:2005mi}. In our computation the leading 
approximation to closed strings production (i.e. the exponential terms) does not 
change at $k=1$. The power-law corrections that determine the behavior of the system 
are model-dependent and would require a more detailed study.

\paragraph{Extended branes}
The result found for long string emission 
seems peculiar to the D0-brane since it comes from the exponentially large contribution 
of states with large radial momentum (of order $e^{2\pi P/k}$)
to the $\slc$ part of the amplitude~(\ref{prodlongexpl}), see eqn.~(\ref{longbehav}). To obtain 
a \textsc{uv} finite long strings emission one may try to start with extended D-branes in AdS$_3$ (the 
analogue of an \textsc{fzzt} brane of Liouville theory).

We consider a symmetry-breaking D2-brane of AdS$_3$ in type IIB superstrings, 
made out of a D2-brane of the cigar. This D2-brane covers the whole 
AdS$_3$ space-time and has a magnetic field on its worldvolume; the latter is parameterized by 
a (quantized) label \mbox{$\sigma \in [0,(1+\frac{1}{k})\frac{\pi}{2})$}~\cite{Ribault:2003ss,
Israel:2004jt}.\footnote{The boundary \textsc{cft} 
for the D2-brane in $\slc$ is not completely well understood; see 
in particular~\cite{Eguchi:2003ik,Fotopoulos:2004ut} for other proposals. We follow 
here the approach of~\cite{Ribault:2003ss,Israel:2004jt}, continued to 
integer $k$~\cite{Israel:2005fn}. Note that qualitatively the features 
discussed in this paragraph don't depend on the particular D2-brane boundary 
state chosen. } The modulus squared of the one-point function for 
the D2-brane in $\slc$  scales asymptotically with the radial momentum as  
\mbox{$\exp \{\frac{2\pi P}{k} (\frac{2k\sigma}{\pi}-k-1)\}$}. Therefore, if one adds the rolling tachyon 
boundary deformation~(\ref{defsuper}) to the AdS$_3$ worldsheet \textsc{cft} with a D2-brane boundary, 
one finds that long string radiation is exponentially 
suppressed at high energies.

However, the physical interpretation of this worldsheet boundary \textsc{cft} 
is very different from the D0-brane example. In the open string sector of the 
D2-brane, the identity (i.e. the \textsc{ns} open string vacuum) is not normalizable. 
The physical open string tachyon on the extended brane 
belongs to the continuous representations of $\slr$,with $j=1/2+iP$, $P\in \mathbb{R}_+$ and its 
wave-function is delta-function normalizable. Therefore one cannot 
interpret the worldsheet theory with the boundary deformation~(\ref{defsuper}) as the decay of the unstable 
D2-brane. The latter would correspond to a boundary deformation built on continuous representations of $\slr$. 
We leave as an open problem the analysis 
of this more complicated worldsheet boundary \textsc{cft}.

\subsubsection{Short strings production} The closed string spectrum  
in AdS$_3 $ contains also discrete representations in the 
range~(\ref{spinbound}). They correspond to ``short strings'' 
trapped inside AdS$_3$~\cite{Maldacena:2000hw}. 
By following the same steps as above, we arrive to the 
following expression for the  imaginary part of the 
annulus amplitude, that is interpreted as the mean number 
of emitted physical short strings: 
\begin{align}
\bar{\mathcal{N}}_d  = \text{Im} \, \mathcal{A}_d = 
\frac{1}{2k\prod_i R_i} \sum_{|\phi \rangle \in \mathcal{H}_d} \sum_{2\tilde \jmath =1}^{k-2} 
\sin \left( \frac{\pi (2\tilde
    \jmath +1)}{k} \right)
 \int_{\frac{1}{2}}^{\frac{k+1}{2}}\!\!\!\! 
\di j\ \sum_{\ell, w \in \zi}  \sin \frac{\pi (2j-1)}{k}
 \ \times
 \nonumber \\  \times \ 
 \sum_{a,b} \sum_N (-)^{bF} D(N)\ 
\frac{(-)^a}{|2j-1+ kw| \sinh^2 \left( \frac{\pi E_a}{\sqrt{2k}} 
+ \frac{i\pi a}{2} \right)}\ \times \nonumber \\ 
 \times 
\delta \left(j-\frac{1}{2}+\frac{k}{2}w 
-\sqrt{\frac{1}{4}+k\left[ N+
\frac{\tilde \jmath(\tilde \jmath+1)}{k}  + \frac{\sum_i(\frac{n_i}{R_i})^2}{2}
-w (\ell + \frac{a}{2})
 -\frac{w +(1-a)}{2} \right]} \right) ,
\label{shortem}
\end{align}
summing over physical states belonging to the sub-Hilbert space $\mathcal{H}_d$ of 
closed strings made with discrete $\slr$ representations. 
The on-shell spacetime energy is now
\begin{equation}
E_a = 2(j+\ell + \frac{a}{2})+ kw
\label{disceneg}
\end{equation}
where $j$ solves the delta-function constraint in eqn.~(\ref{shortem}). 

One of the main 
characteristics of the emitted short strings is that 
they don't go away far from the locus of the brane. 
Classical short strings with $w=0$ are time-like geodesics with a periodic motion
around the origin, while  classical spectral-flowed 
short strings have a periodic breath mode~\cite{Maldacena:2000hw}:
\begin{equation}
e^{i\phi}\sinh \rho = ie^{iw\sigma} \sinh \rho_0 
\sin \frac{2j \tau}{k} \quad \quad \text{with}\quad \cosh \rho_0 = 1+\frac{\ell}{j}\, ,
\end{equation}
where $(\tau,\sigma)$ are the worldsheet coordinates and $(\rho$, $\phi)$ the 
space-like global coordinates in AdS$_3$, see eqn.~(\ref{globalcord}). 
Unlike the continuous representations, there are on-shell discrete states in the sector $w=0$. 
The contribution of short strings with $w=0$ is \textsc{uv} finite since the upper 
bound on $j$ of eqn.~(\ref{spinbound}) implies an upper bound on the oscillator number $N$ 
in this sector.

Let us now examine the high energy behavior of the amplitude~(\ref{shortem}) in 
the sectors of non-zero spectral flow $w$. As for long strings,  
the density of states grows exponentially with the oscillator number $N$ while the amplitude is 
exponentially suppressed as a function of the energy~(\ref{disceneg}). Because of 
the bounds~(\ref{spinbound}) on the $\slr$ spin $j$, for fixed large spectral flow $w$, 
only the discrete states with $\ell \sim \nicefrac{N}{w}-\nicefrac{kw}{4}$ will contribute for $N \gg 1$. 
Therefore, from eqn.~(\ref{disceneg}) giving the space-time energy, 
the amplitude is weighted at large $N$ by a factor  
\begin{equation}
\frac{1}{\sinh^2 \frac{\pi E}{\sqrt{2k}}}
\sim e^{-\pi\sqrt{\frac{k}{2}} w - 2\pi \sqrt{\frac{2}{k}} \frac{N}{w}} \, ,
\label{shortbehav}
\end{equation} 
similar to what we obtained above for long strings, see
eqn.~(\ref{behavintp}). It is interesting that we obtain, 
for short strings in a given spectral flow sector, the field theory entropy 
$ S(E) \sim \sqrt{E}$ expected from AdS$_3$/\textsc{cft}$_2$ duality~\cite{Maldacena:1997re} 
using the bound~(\ref{spinbound}) and the on-shell condition. 

The leading contribution to the asymptotic density of states for the 
transverse degrees of freedom, with a reasoning 
similar to the continuous representations, is given by~(\ref{cardy})
with the same $c_\text{eff} = 12 -\nicefrac{6}{k}$.\footnote{One can also find an upper bound on the 
degeneracy of states using the explicit expansion of the discrete representations characters, 
confirming this result.} In contrast with the long strings sector, 
the contribution of the square of  the $\slc$ one-point function coefficient 
to the annulus amplitude, see eqn.~(\ref{shortem}), 
is of order one. The net effect is that the exponential suppression of short 
strings emission wins over the asymptotic 
density of states at high energy. Therefore the number of emitted short strings, 
evaluated at tree level, is  \textsc{uv} finite. It means that, for a given 
AdS$_3$ radius (i.e. fixed $k$), one can choose the string 
coupling constant $g_6$ small enough in order to get an arbitrarily large fraction of the energy 
dissipated into long strings, rather than into short strings.

\paragraph{Infrared divergence}
Because the spectrum of discrete $\slr$ representations contains states with zero space-time 
energy, there is an infrared divergence in the computation of the 
mean number of emitted short strings. From the worldsheet \textsc{cft} 
point of view it comes from the extra primaries at $E=0$, that 
we discussed briefly in subsection~\ref{rollflat}. The only $\slr$ physical states 
with zero energy (and compatible with the \textsc{gso} projection) 
are the $\cn =2$ Liouville  interaction~\cite{Hori:2001ax} and 
its image under one unit of spectral flow. It is also worthwhile to 
remind that this divergence is eliminated 
if we choose the ``full-S-brane'' solution~(\ref{rolbounddef}) instead of the 
``half-S-brane''~\cite{Lambert:2003zr}. There is no indication that 
this infrared divergence is more harmful that in flat space-time.

\paragraph{Flat space-time limit}
It is interesting to have a look at the $k\to \infty$  limit, 
in order to connect our results to those obtained in flat space-time.\footnote{We 
thank Nissan Itzhaki for discussions about this issue.} Let's consider first 
the continuous $\slr$ representations. Unflowed states ($w=0$) are not part of the physical 
spectrum even for $k=\infty$ as they correspond to space-like geodesics. Long strings 
decouple in the flat space-time limit, because their mass is of order $kw$. The unflowed discrete  
representations play the main role in the $k\to \infty$ limit (the contributions of 
short strings with $w\neq 0$ to the amplitude are also suppressed in the large $k$ limit). 
Indeed the upper bound on 
the $\slr$ spin $j$, see eqn.~(\ref{spinbound}), which prevents them from giving a significant contribution 
to closed string emission, disappears. At large $N$ states with $j \sim \sqrt{kN}$ will give 
contributions to the amplitude~(\ref{shortem}) 
for closed string emission of order $e^{-2\pi \sqrt{2N}}$, like 
in flat space-time. As the density of states~(\ref{cardy}) converges also to the flat space-time value 
in the $k\to \infty$ limit, we recover the main aspects 
of the flat space-time results found in~\cite{Gaiotto:2003rm,Lambert:2003zr}. In order to get 
a precise matching one needs to be very careful about the order of limits.
The energy dissipated into short strings is independent of $g_6$ at leading order. 
For fixed $g_6$ if one increases 
$k$, i.e. decreases the AdS$_3$ curvature, one expects to reach a turnover point above which the short 
strings production is dominant.

\paragraph{Summary of closed string emission}
Our computation of the annulus amplitude shows that, at small string coupling, 
almost all the energy radiated into closed strings resides
in long strings with large winding number. The short strings production 
is finite and carries only a small fixed fraction of the energy. 
The radiation of long strings changes effectively the string coupling
constant (in the interior of the shell of closed strings that are emitted), 
since $g_s^2 \propto \nicefrac{1}{Q_1}$, 
where $Q_1$ is the number of fundamental strings that build up the background. 
This quantity will of course receives higher order perturbative corrections, e.g. 
multi-closed strings emission which is of order $g_s^{2(N-1)}$ for $N$-particle 
states.

In flat space-time, the characteristics of the closed string radiation led 
Sen to the {\it open string completeness conjecture}~\cite{Sen:2003iv},
which states that the open string field theory description 
gives a complete description of the decay, in particular  ``takes 
care'' of the apparently large back reaction due to massive 
closed strings. This open string description is approximated 
by the worldvolume Born-Infeld-like action for a non-\textsc{bps} D-brane, 
which exhibits  the distinctive features 
of tachyon condensation as described by 
closed string radiation, namely pressureless matter in the asymptotic 
future and no plane waves excitations near the
minimum of the potential, signaling the presence of ``tachyon dust''  made 
of  non-relativistic massive closed strings.

In the example studied here, D-particle decay in AdS$_3$, 
the radiation is made of macroscopic long strings. 
Both cases are attempts to take perturbation theory beyond its range of validity. 
The two pictures, flat space-time and AdS, coincide in the region 
where perturbation theory is valid, i.e. for a timescale of order $\sqrt{\alpha'}$,
before the long strings radius become significantly larger than the string scale, 
and the two pictures diverge after that time.

\subsection{Comparison with non-critical strings} 
As we demonstrated above, in AdS$_3$ the high-energy divergence in closed 
strings emission from the brane decay is regularized non-perturbatively. 
It has been suggested in~\cite{Karczmarek:2003xm} that for brane decay in non-critical strings 
the closed string emission is \textsc{uv}-finite a tree level, because the high-energy density of states is 
somewhat lower than in flat space-time.  As anticipated in the introduction it is very unlikely that the 
high-energy behavior of non-critical strings could be softer than string theory in anti-de Sitter space-time. 
We would like to clarify this issue by uncovering similar divergences in non-critical strings (to those in 
AdS$_3$), although the mechanism of closed string emission is different.

For concreteness we consider unstable D-branes in superstring theories of the form $\mathbb{R}^{2n,1} \times \mathbb{R}_Q$. In order to obtain a stable supersymmetric background, and regularize the strong coupling region due to the linear dilaton, one has to compactify one of the spatial directions at a precise radius 
and add an $\cn = 2$ Liouville potential. Then 
one obtains the worldsheet \textsc{cft} $\mathbb{R}^{2n-1,1} \times \slc |_{2/(4-n)}$ with the appropriate \textsc{gso} 
projection. 

As in~\cite{Karczmarek:2003xm} we could consider first a non-\textsc{bps} brane extended along the linear dilaton direction $\rho$, i.e. made of a D1- or D2- brane of the cigar.\footnote{The analogous problem 
of the decay of an \textsc{fzzt} brane in 2D bosonic string theory has been discussed in~\cite{McGreevy:2003ep}.} However, as for the D2-brane in AdS$_3$ 
discussed above, the rolling tachyon boundary deformation~(\ref{defsuper}) is 
built with the identity of the $\slc$ coset. The latter is not normalizable on the extended brane because of the 
asymptotic linear dilaton. It means that, while being a consistent worldsheet
boundary \textsc{cft}, it does not represent the decay of the physical open 
string tachyon living on the brane. 

The open string tachyon on the extended brane 
belongs to the continuous representations of $\slc$, as for the extended brane in AdS$_3$ discussed above. 
To analyze its decay using \textsc{bcft} 
methods one should consider instead the worldsheet theory deformed by a boundary marginal 
deformation of the asymptotic form 
\begin{equation}
\delta S = \lambda\oint_{\p \Sigma} \di \ell \ G_{-1/2} \ e^{-
\frac{1/2+iP}{\sqrt{2k}}\rho + 
\sqrt{\frac14 -\frac{P^2+1/4}{2k}} X^0}\, \sigma^1 \, .
\end{equation}
As in AdS we do not know how to solve the theory in the presence of this boundary term in the 
action even in the case $P=0$, i.e.  homogeneous decay. 
It would be interesting also to understand the physics of the rolling tachyon~(\ref{defsuper}) in 
the non-critical strings context, in particular to understand if and why closed strings emission is finite there. In bosonic strings, similar statements can be made, considering an \textsc{fzzt} extended brane~\cite{Fateev:2000ik} of 
Liouville theory.\footnote{In both cases, bosonic and supersymmetric, the bulk Liouville potential is not important 
in the discussion if one takes the boundary cosmological constant much larger than the bulk cosmological constant, 
in such a way that the brane "dissolves" in the weak coupling region.}

The exact non-critical string analogue of the unstable D-particle that
we study in AdS$_3$ is to  consider a non-\textsc{bps} D0-brane made with a D0-brane of the 
$\slc$ coset, i.e. localized at the tip of the cigar (similar to a 
\textsc{zz} brane of bosonic Liouville theory). In the open string sector the identity of the 
$\slc$ \textsc{cft} is normalizable therefore one can describe the condensation of the 
physical open string tachyon with the boundary deformation~(\ref{defsuper}). 
Let's consider six-dimensional superstrings, i.e. 
$\mathbb{R}^{5,1} \times \slc|_{k=2}$ as an example.
The one-point function for an operator in the the continuous representations of the coset is:
\begin{equation}
|\langle V^{\mathfrak{sl}_2/\mathfrak{u}_1}_{j=1/2+iP\, s \bar s}
\ e^{iEX^0}\rangle|^2 
\propto \delta_{s,-\bar s} \ \frac{\sinh 2\pi P \sinh \pi P}{\cosh 2\pi P + \cos \pi s} \frac{1}{\sinh^2 \pi E} \, ,
\end{equation}
where $(s,\bar s)$ are the $\zi_4$-valued left and right momenta of $U(1)_2$. 
In the non-critical superstring the on-shell condition is simply
\begin{equation}
E = \sqrt{P^2+1/4+\frac{s^2}{2} + \mathbf{p}^2+2N-1}\, ,
\end{equation}
where $P$ is the radial momentum in $\slc$ and $\mathbf{p}$ the $\mathbb{R}^5$ momentum transverse 
to the brane.
The density of states that appears in the computation of the annulus amplitude, including 
the power-law corrections,  
is the same as our $\slr |_{k=2} \times S^1 \times \mathbb{R}^4$  previous example: 
\begin{equation}
D(N) \sim N^{-2} e^{2\pi \sqrt{\frac{3N}{2}}}\, .
\end{equation}
So at large $N$ and $P$ the mean energy emitted by the brane decay scales like 
\begin{multline}
\bar{E} \sim \int \di^5 \mathbf{p} \int \di P \sum_N \frac{1}{N^{2}}\, 
 e^{2\pi \sqrt{\frac{3N}{2}}}\ 
e^{\pi P}\ e^{-2\pi \sqrt{P^2+2N+\mathbf{p}^2}}\\ 
\sim  \int \di P \sum_N \frac{1}{N^{2}}\, 
 e^{2\pi \sqrt{\frac{3N}{2}}}\ 
e^{-\pi (2\sqrt{P^2+2N}-P)} 
\left[ \int \di \rho \  e^{-\frac{\pi \rho^2}{\sqrt{P^2+2N}}} \right]^5 \, .\\ 
\end{multline}
One can use a saddle point approximation to evaluate the integral over
$P$ at large $N$. As for the D0-branes in AdS$_3$, the exponential terms in $\sqrt{N}$ cancel, leaving 
the power-law corrections that determine the behavior of the amplitude. 
One obtains that the total energy radiated into closed strings, evaluated at tree level, 
diverges like $\sum_N N^{-1/2}$. Therefore in non-critical superstrings the \textsc{uv} divergence 
of closed strings emission remains.\footnote{The decay 
of \textsc{zz} branes in 2D bosonic string theory has been considered in~\cite{Klebanov:2003km}, 
and recently extended in~\cite{He:2006bm} to higher dimensions. In all those cases the 
closed strings emission is also divergent.}

\subsection{Open string pair production}
\label{openpair}
One possible way of radiating the energy of the D-brane, 
as an intermediate stage of the brane decay, is to consider 
open string pair creation. It is not completely clear what
is the meaning of this process since the open strings
loose their support. However if this emission happens to be
divergent it singles the breakdown of open string perturbation
theory. This would render unreliable both the open string 
and the closed string computations at tree level.

Let's consider the boundary theory of an unstable D-particle in the presence 
of the boundary deformation~(\ref{halfrolbounddef}). At past 
infinity $x^0 \to - \infty$ this interaction vanishes and the spectrum 
of open strings can be read from the D-particle 
partition function~(\ref{dpartannulus}). A very important difference 
with the closed string spectrum is that the $\slr$ contribution splits 
into an $\slc$ part and a U(1) temporal part whose  
zero-modes are independent from each other. Therefore the scaling 
of the energy with the oscillator number is the same as in flat 
space: $E \sim \sqrt{N/\alpha'}$. However, despite the fact that 
the open string spectrum contains the identity, the 
effective central charge entering in the asymptotic density of states~(\ref{cardy}) 
is the same as in the closed string sector ($c_\text{eff}=12-\nicefrac{6}{k}$), 
because this leading contribution is evaluated by 
modular transform of the annulus amplitude to the closed string channel where
the minimal conformal dimension in the $\slc$ \textsc{cft} is $\Delta_\text{min} =\nicefrac{1}{4k}$. 

To conclude, AdS$_3$ string theory looses in the tree-level approximation 
one of  its prominent features in the open string sector of the unstable D-particle. 
It has an Hagedorn growth of the density of states at high energies, with the same Hagedorn 
temperature as a non-critical superstring with 
a dilaton slope $Q =\sqrt{\nicefrac{2}{k}}$.\footnote{Of course since this D-brane is
unstable the thermodynamics of open strings attached to it is not well-defined.}
This has to be contrasted with the spectrum of  
AdS$_2$ branes~\cite{Lee:2001gh}. This fact may probably not be true
non-perturbatively since string theory on AdS$_3$ is dual to a field theory
and should not exhibit an Hagedorn growth of the density of states in any sector.
In any case we will show that even with this Hagedorn behavior 
there is a finite average number of produced open strings because the Hagedorn 
temperature is higher than is flat space.

The computation of the amplitude for pair production of open strings uses 
the boundary two-point function, which plays the role of a Bogolioubov 
coefficient~\cite{Gutperle:2003xf}. We have already seen that 
it is related to the energy density appearing 
in the open string annulus amplitude of eqn.~(\ref{tachpart}); we shall 
now discuss how to obtain its actual expression using the orbifold 
construction of subsection~\ref{subsecorbicons}.

It is quite interesting that the open string annulus amplitude for the 
rolling tachyon, eqn.~(\ref{spectraldensads}), that we obtained by channel duality from 
the closed string annulus amplitude~(\ref{closechanrolads}), does not exhibit poles 
corresponding to on-shell open strings pair-produced by the time-dependent interaction. 
As we saw above its imaginary part signals the production of closed strings. 
Some comments about this issue can be found in~\cite{Karczmarek:2003xm}.

We consider in the following the bosonic case for technical convenience.  
The extension to the superstring, using the results of~\cite{Fukuda:2002bv} 
is possible. We start with the 
boundary two-point function in ordinary Liouville theory (with $Q=b+b^{-1}$). 
The boundary two-point function, or reflection amplitude, 
for an open string "tachyon" of dimension $\Delta = \alpha (Q-\alpha)$, interpolating between the boundary 
conditions $s_1$ and $s_2$, is given by~\cite{Fateev:2000ik}:  
\begin{multline}
d_b (\alpha|s_1,s_2) = \frac{b}{2\pi}\left( 
\mu \pi \frac{\Gamma(b^2)}{\Gamma(1-b^2)} 
\right)^{\frac{Q-2\alpha}{2b}}
\Gamma \left(\frac{2\alpha}{b}-\frac{1}{b^2}\right) 
\Gamma \left( 2b \alpha - b^2- 1\right) \times \\
\times \frac{S^{(0)}_b (2\alpha)}{S^{(0)}_b (\alpha + \frac{s_1+s_2}{2i})
S^{(0)}_b (\alpha - \frac{s_1+s_2}{2i})S^{(0)}_b (\alpha + \frac{s_1-s_2}{2i})
S^{(0)}_b (\alpha - \frac{s_1-s_2}{2i})} \, , 
\end{multline}
using the special function $S^{(0)}_b (x)$ defined by eqn.~(\ref{Sfct}). First,  
as discussed in subsection~\ref{rollflat}, we are interested in the limit of 
vanishing Liouville interaction $\mu \int \di^2 z \exp{2b\rho (z,\bar z)}$,  
while keeping fixed the boundary cosmological constant $\tilde \lambda$, i.e. 
with a fixed boundary interaction of the form:
\begin{equation}
\delta S =  \frac{\tilde \lambda}{2} \oint_{\p \Sigma} \di \ell \ e^{b \rho (\ell )}\, .
\label{boundpotliouv}
\end{equation}
This regime corresponds to a brane which "dissolves" way before the Liouville potential becomes 
important because the boundary potential~(\ref{boundpotliouv}) acts like a barrier for open string modes. 
The boundary parameter $s$ is related to the boundary cosmological constant as 
\begin{equation}
\tilde \lambda^2 = \frac{4\mu}{\sin \pi b^2} \cosh^2 (\pi b s) \, ,  
\end{equation}
such that we consider the limit $\mu \to 0$, $s_{1,2}\to \infty$, with 
$(s_1-s_2)$ fixed. 
As in~\cite{Gutperle:2003xf}, using the asymptotics of $S^{(0)}_b (x)$ 
one gets for the boundary two-point function:
\begin{multline}
d_b (\alpha|s_1,s_2) = 
\frac{b}{2\pi}\left( 
\frac{\pi^2 \tilde \lambda_1 \tilde \lambda_2}{\Gamma (1-b^2)^2}
\right)^{\frac{Q-2\alpha}{2b}}
\Gamma \left(\frac{2\alpha}{b}-\frac{1}{b^2}\right) 
\Gamma \left( 2b \alpha - b^2- 1\right) \times \\
\times \frac{S^{(0)}_b (2\alpha)}{S^{(0)}_b (\alpha + \frac{s_1-s_2}{2i})
S^{(0)}_b (\alpha - \frac{s_1-s_2}{2i})} \, . 
\label{boundtbl}
\end{multline}
The boundary two-point function for time-like boundary 
Liouville theory is obtained in the limit $b\to i$. However in this limit 
the function $S^{(0)}_b (x)$ has an infinite number of poles accumulating for 
every $x \in i \zi$. One therefore should give some prescription for this 
limit; we refer the reader to~\cite{Gutperle:2003xf,Fredenhagen:2004cj} for details 
about these issues. 

In the orbifold construction of the rolling tachyon in 
AdS$_3$, see subsection~\ref{subsecorbicons}, we start with branes on the 
covering space of the orbifold, i.e. $\slc^\textsc{v}_\infty \times \mathbb{R}^{0,1}$, 
and sum over the images under the geometric identification. In the orbifold theory 
there are new boundary operators corresponding to open strings stretched between 
the brane and one of its images.\footnote{See the reference~\cite{Matsubara:2001hz} for 
a detailed analysis of the rational \textsc{cft} analogue.} 

Open strings in the rolling 
tachyon \textsc{bcft} stretched between a brane and its image under the translation
$\mathfrak{T}^r$ correspond to boundary operators interpolating between 
the  boundary cosmological constants\footnote{There is a factor of $\sqrt{2}$ in the bosonic
theory compared to the superconformal case discussed in subsection~\ref{subsecorbicons}.} 
$\tilde \lambda_1$ and $\tilde \lambda_2 = \tilde \lambda_1 e^{2\pi r/\sqrt{k}}$, 
i.e. the temporal part of the boundary two point function is 
given by the $b\to i$ limit of~(\ref{boundtbl}) with $s_1-s_2 = 2i r/\sqrt{k}$. 
The annulus amplitude for open strings stretched between two such branes contains only the coset 
character $ch_\mathbb{I} (r; \tau)$ coming from the identity representation of 
$\widehat{\mathfrak{sl}}(2,\mathbb{R})$, see the open string partition 
function~(\ref{spectraldensads}).

Using the prescription of~\cite{Gutperle:2003xf} one 
finds that the temporal part of the boundary two-point function 
in AdS$_3$, for an open string 
vertex operator of the form 
$e^{i\upsilon X^0}$ of conformal dimension $\Delta= -\upsilon^2$ 
is given by:\footnote{The methods developed in~\cite{Fredenhagen:2004cj} are 
more appropriate to the Euclidean $c=1$  Liouville theory.}
\begin{multline}
d_i (\upsilon|r) = \left( \pi \tilde{\lambda} e^{\frac{\pi r}{\sqrt{k}}} 
\right)^{2i\upsilon} \ \frac{S_1^{(0)} (1-2i\upsilon)}{S_1^{(0)} \left(1-i\upsilon-  \frac{ir}{\sqrt{k}}\right)
S_1^{(0)} \left( 1-i\upsilon+  \frac{ir}{\sqrt{k}} \right)
} \ \times\\
\times \ 
\frac{\sinh \pi (\upsilon+  \frac{r}{\sqrt{k}})
\sinh \pi (\upsilon-  \frac{r}{\sqrt{k}})}{\sinh^2 2\pi \upsilon}\, .
\label{sfctpart}
\end{multline}
This result is as expected closely related 
to the (bosonic version of the) energy density~(\ref{tachpart}). Using 
the identity $S_b^{(0)} (x) S_b^{(0)} (Q_b -x)=1$ we observe that   
the first line of the boundary two-point function~(\ref{sfctpart}) is a pure phase.

The $\slc$ part of the boundary operator belongs to the coset module built on the 
$\slr$ operator in the identity representation with $J^3 = r$, see the 
orbifold action~(\ref{translact})
and the partition function~(\ref{spectraldensads}), and can be normalized to one. 
Therefore the modulus of the boundary two-point function for the rolling tachyon in AdS$_3$ in the $r$-sector is:
\begin{equation}
\left| d(\upsilon|r)_\text{AdS} \right| = \left| \frac{\sinh \pi (\upsilon+  \frac{r}{\sqrt{k}})
\sinh \pi (\upsilon-  \frac{r}{\sqrt{k}})}{\sinh^2 2\pi \upsilon}\right|\, .
\label{b2pf}
\end{equation}

Following~\cite{Gutperle:2003xf} one can identify the boundary reflection amplitude 
$d(\upsilon|r)$ with the Bogolioubov coefficient $-\gamma_\upsilon^{\text{out}\, *} =
\beta_\upsilon / \alpha_\upsilon$. Then the vacuum amplitude giving the rate of pair production 
(see~\cite{Karczmarek:2003xm} for details) is obtained as:
\begin{align}
\mathcal{W} &=-\text{Re}\, \ln \langle \text{out} | \text{in} \rangle = -\frac{1}{4} \sum_{N,r} D(N,r)
\ln  (1 - |\gamma_\upsilon(N,r)|^2)    \nonumber\\
&= -\frac{1}{4} \sum_{N,r} D(N,r) \ln  \left(1 - \left|d(\upsilon(N,r)|r)_\text{AdS}\right|^2 \right)\, .
\end{align}
The on-shell energy  for an open string in the $r$-sector is:
\begin{equation}
\upsilon (N,r) = \sqrt{\frac{r^2}{k}+r+N-1 }
\end{equation}
Using the asymptotics of~(\ref{b2pf}) for large $\upsilon$ and the high energy density of 
open string states as discussed above, we find the \textsc{uv} asymptotic behavior
\begin{equation}
\mathcal{W}\sim \sum_{r \in \zi} \sum_{N} e^{-4\pi
\sqrt{\frac{r^2}{k}+r+N-1}\left(1-\sqrt{1-\frac{1}{4k}}\right)}\, .
\end{equation}

Therefore open string pair production from the D0-brane decay in AdS$_3$ is 
exponentially convergent at high 
energies because, as in~\cite{Karczmarek:2003xm} for non-critical strings, the high-energy 
density of states is lowered compared to flat space-time. 
The same conclusion holds for the superstring case. It shows that open string perturbation 
is not invalidated by an infinite open string pair production.  The average number of 
emitted open string pairs  does not 
depend on the string coupling $g_s$ in the leading order,   thus open string pair 
production is negligible compared to the emission of closed long strings 
at weak coupling. In the  $k\to +\infty$ flat space-time limit, the open string pair production 
is governed by power-law corrections and its divergences 
depend on the dimensionality of the brane and the specific energy moment considered.

\section{Comments on holography}
\label{secholog}
One of the main reasons to study string theory on AdS$_3$ is that, 
besides being an interesting example of curved space-time, it is an incarnation 
of the AdS/\textsc{cft} holographic correspondence between gravity and 
field theory~\cite{Maldacena:1997re}. It is actually the only example at our disposal for which 
the string theory is under control, at least in the perturbative regime, 
because of its realization as an $\slr$ \textsc{wzw} model. 

However life is not so simple; 
the space-time two-dimensional \textsc{cft} dual to AdS$_3$ 
backgrounds with only \textsc{ns-ns} flux (i.e. with a \textsc{wzw} model
description) is {\it singular} because the brane configuration that 
realizes this field theory on their worldvolume in the infrared limit 
can fragment at no cost of energy~\cite{Seiberg:1999xz}. One can move away from  this singularity 
by turning on moduli in the space-time \textsc{cft}. They correspond in 
space-time to Ramond-Ramond fluxes that invalidate
the \textsc{rns} construction of the worldsheet theory. As these fluxes prevent  
long strings from expanding towards the boundary, it modifies quite drastically 
the physics of the decay. We refer the reader to 
the review~\cite{David:2002wn} for more details and references.

Once focusing on the \textsc{bps} sector of the space-time \textsc{cft} one can 
argue that certain quantities are protected by supersymmetry, allowing to
compare results from the supergravity/worldsheet \textsc{cft} side and from the space-time \textsc{cft} at  
a non-singular point in the moduli 
space~\cite{deBoer:1998ip,Aharony:1999ti,Argurio:2000tb}. 
We are interested here  in the decay of a non-\textsc{bps} state, materialized on the gravity 
side as a D-particle, therefore there is no simple way to 
find its dual in the space-time theory. We will collect in this section 
some facts that may hint  the solution of this problem.

\paragraph{The D1-D5 system and the symmetric product CFT}
Among the various  AdS$_3$  backgrounds, type \textsc{iib} 
on AdS$_3 \times$ S$^3 \times$ T$^4$ has the most studied 
holographic dual; indeed it is related to the microscopic construction 
of supersymmetric black holes. This space-time is obtained 
as the near-horizon limit of $k$ NS5-branes wrapped on T$^4$ and $Q_1$ fundamental 
strings smeared on the compact manifold.

In the S-dual picture, i.e. as a collection of D1- and D5-branes, the open string 
massless degrees of freedom give a two-dimensional $U(k) \times U(Q_1)$ quiver gauge theory with 
$\mathcal{N}=(4,4)$ supersymmetry. At very low energies, the Higgs
branch of the quiver theory flows to an $\mathcal{N}=(4,4)$ superconformal theory, a non-linear 
sigma model for the hypermultiplets whose target space is an hyper-K\"ahler manifold, 
together with a free $\mathcal{N}=(4,4)$ \textsc{cft} corresponding to the center of 
mass coordinates on T$^4$. The target space of the 
former can be seen as the moduli space $\mathcal{M}$ of $Q_1$ instantons in a $U(k)$ 
gauge theory on T$^4$. This \textsc{scft} lies  in the moduli space of the symmetric product
$(T^4)^{kQ_1}/S(kQ_1)$ and has a central charge $c=6kQ_1$~\cite{Dijkgraaf:1998gf}. 
For the ``pure'' D1/D5-system --~or equivalently via S-duality the 
NS5/F1 background without Ramond-Ramond fluxes~-- this conformal field theory 
is singular, because of the small instanton singularities in $\mathcal{M}$. From the 1+1
gauge theory perspective it is obtained by turning off the Fayet-Iliopoulos (\textsc{fi})
terms and the theta-angle. This process corresponds to a situation where a long D-string, viewed as an
instanton in the D5-brane worldvolume, ``leaves'' the system of branes~\cite{Seiberg:1999xz}. 
The singularity occurs where classically the Higgs branch 
and the Coulomb branch meet.

\boldmath
\subsection{Instantons, sphalerons in AdS$_3$ and their holographic dual}
\unboldmath
Before addressing the difficult problem of finding the dual of the unstable 
D-particle, we would like to understand precisely the meaning of the 
AdS$_3$ D-instanton constructed in section~\ref{secbraneads} in the context of the space-time 
\textsc{cft}. These two objects are closely related for  
two reasons. 

First, the D-instanton and the D-particle boundary states in AdS$_3$
differ only by the gluing conditions along the time direction; they come
from the same D-brane in the coset $\slc$. 
Moreover,  the worldsheet analysis of the D0-brane decay 
involves also D-instantons in an interesting way.  
Following~\cite{Lambert:2003zr}, we can interpret the imaginary 
part of the annulus diagram for the rolling tachyon, eqn.~(\ref{imannul}), as a sum over 
contributions from an array of D-instantons (whose annulus amplitude is 
given in~\cite{Israel:2005ek}) along the imaginary time axis.

Second, and more importantly, it has been 
argued~\cite{Harvey:2000qu} that the unstable D0-brane  
in type \textsc{iib} superstring theory is a sphaleron, 
i.e. an unstable classical solution associated with a non-contractible loop 
in configuration space~\cite{Manton:1983nd}. 
Inspired by this idea, it has been shown that in AdS$_5 \times$ S$^5$ 
the D-particle located at the origin of global coordinates in AdS$_5$ 
is indeed a sphaleron in the gauge theory~\cite{Drukker:2000wx}. 
The latter is related to the \textsc{sym} instanton in the sense that it is a 
classical solution of theory at the maximum of the potential barrier between 
two vacua of the gauge theory for which the instanton represents the tunneling
process. It allowed to argue that the sphaleron survives at strong 
coupling --~which is the regime where the holographic duality can be probed on
the string side~-- because the associated instanton is a topologically 
stable and \textsc{bps} object.

\paragraph{Instantons in the D1/D5 system}
As discussed above the worldvolume theory of the D1/D5 system is described in the infrared 
limit, dual to string theory on \mbox{AdS$_3 \times$ S$^3 \times$ T$^4$}, 
by a non-linear sigma model on $\mathcal{M}$ with $\mathcal{N}=(4,4)$ superconformal 
symmetry. Near the point where one D1-brane could leave the system (for which a Coulomb 
branch opens up) the dynamics of the system is well described by a non-linear sigma 
model on the cotangent bundle of $\mathbb{CP}^{k}$. Then one gets instantons 
associated with maps $\mathbb{C} \to \mathbb{CP}^{k}$ described as 
vortices in the linear sigma-model description of the theory~\cite{Witten:1993yc} 
(see~\cite{Chen:2006ps,Rey:2006bz} for recent works 
on this subject). The action  
of these \textsc{bps} instantons vanishes  
for zero \textsc{fi} terms, i.e. when the singularities are blown-down, and zero theta-angle. 
The AdS/\textsc{cft} dictionary matches the \textsc{fi} terms with the 
self-dual part of the \textsc{ns-ns} two-form on T$^4$, in the D1/D5 description. 
One may identify these instantons with the Euclidean 
D1-branes wrapping holomorphic cycles of T$^4$ discussed in~\cite{Mikhailov:1999fd}. 
With an \textsc{ns-ns} two-form 
turned on, a D1-charge is induced in the presence of a non-zero instanton charge density through the coupling 
$\int_{\mathbb{R} \times S^1 \times T^4} B \wedge C_2 \wedge F$ on the D5-brane worldvolume. 

It is however not clear to us what is the dual of the D(-1)brane, that we constructed 
in the AdS$_3 \times$ S$^3 \times$ T$^4$ background with only \textsc{ns-ns} fluxes 
(i.e. at the singular point) since a \mbox{D(-1)-D1-D5} configuration is not 
supersymmetric.\footnote{We would like to 
thank O.~Aharony, A.~Mikhailov, S.~Rey and D.~Tong for discussions about this problem.}  
The D(-1)-brane is expected to form a \textsc{bps} bound state with the D1/D5 system 
"dissolving" in their wordvolume. As for the D0-D2 bound state, the D(-1) turns into
an electric field in the D1-brane worldsheet, inducing a D(-1) charge through the coupling 
$\int_{\mathbb{R}\times S^1} C_0\, F$ in the D-string action. Similarly an Euclidean  D3-brane  
wrapped on the T$^4$ can dissolve in the D5-branes as an electric field.\footnote{In 
the S-dual NS5/F1 description, these two kinds of D-instantons are
both point-like 
in the AdS$_3$ space-time and differ only by the boundary conditions (Dirichlet or Neumann) on T$^4$.} In the 
AdS/\textsc{cft} correspondence the theta angle of the D1/D5 gauge theory 
is identified with a linear combination of the \textsc{rr} axion $C_0$ 
and the \textsc{rr} four-form $C_4$ on T$^4$. Since the theta-angle in two dimensions 
is equivalent to an electric field, it may be related to (a superposition of)  D(-1) and D3 instantons.

An amusing observation about these D-instantons comes from the boundary
worldsheet \textsc{cft} solution itself. To find what kind of object 
the D-instanton represents in the space-time \textsc{cft} it is convenient 
to Fourier transform the worldsheet vertex operators of the Euclidean 
AdS$_3$ (i.e. H$_3^+$) \textsc{cft}, in target Euclidean space-time.  
In this basis the primary operators are written as 
$\Phi^{j} (u,\bar u | z, \bar z)$, where $(u,\bar u)$ are the coordinates 
on the plane where the {\it space-time} \textsc{cft} lives, whereas  
$(z, \bar z)$ are the worldsheet coordinates. Then the one-point 
function in the presence of the D-instanton takes the form~\cite{Ponsot:2001gt}
\begin{equation}
\langle \Phi^{j} (u,\bar u | z, \bar z) \rangle = \frac{1}{|z-\bar
  z|^{2\Delta}} \ \frac{U(j)}{(1+u\bar u)^{2j}} \, .
\end{equation}
The $(z, \bar z)$ dependence on the upper-half plane is fixed by conformal 
symmetry on the worldsheet, while the $(u,\bar u)$ dependence 
is fixed by conformal symmetry on the boundary of space-time where 
the dual space-time \textsc{cft} is defined. It is very 
surprising that this functional form gives exactly a one-point function  
for an operator of dimension $\Delta_{\textsc{st}} = \bar \Delta_{\textsc{st}} = j$ in the space-time \textsc{cft}
on $\mathbb{RP}_2$, i.e. in the presence of a {\it crosscap}.
It leads us to the  rather weird conclusion that, from the point of view of the space-time 
theory, the  D-instanton corresponds to a crosscap.\footnote{One fact supporting 
this interpretation is that the open string sector of the D-instanton 
contains only the vacuum representation of $\slr$ (i.e. of spin $j=0$), 
mapped to an operator of dimension zero in the space-time \textsc{cft}; 
it is consistent with the fact that there are no degrees 
of freedom associated with an orientifold plane.} We will not push forward 
this interpretation because, while  suggested by the 
boundary worldsheet \textsc{cft} computation, its meaning 
is not very clear to us.

As for the D-instantons, the very existence of the non-\textsc{bps} D0-brane 
in the string theory suggests that there is a corresponding sphaleron 
in the space-time \textsc{cft}. It would be very interesting to understand how its decay 
could be related to the holographic dual of long strings emission, i.e. the passage from a 
Higgs to a Coulomb branch  corresponding to the partial fragmentation of the brane stack.


\acknowledgments
We would like to thank C.~Bachas, S.~Elitzur, A.~Giveon, B.~Kol, A.~Mikhailov, S.~Rey, V.~Schomerus, 
D.~Tong, J.~Troost  and S.~Wadia for discussions, 
and especially O.~Aharony, M.~Berkooz and N.~Itzhaki for useful comments on the draft of this paper. 
This work is supported by a European Union 
Marie Curie Intra-European Fellowship under the contract MEIF-CT-2005-024072,
a European Union Marie Curie Research Training Network 
under the contract MRTN-CT-2004-512194, a European Excellence Grant MEXT-CT-2003-509661, 
the American-Israel Bi-National Science 
Foundation, the Israel Science Foundation, the Einstein Center in the Hebrew University 
and by a grant of DIP (H.52). D.I. would like to thank the
Asia-Pacific Center for Theoretical Physics for its hospitality 
during the focus program {\it ``Liouville, Integrability and Branes (3)''}, where part
of this work was done.


\appendix
\boldmath
\section{AdS$_3$ conformal field theory and characters}
\unboldmath
\label{appeads}
In this appendix we recall some facts about string theory on AdS$_3$
that are used in the body of the text. We are interested in the $\slr$ super-\textsc{wzw}
at level $k$, made with a purely bosonic \textsc{wzw} model at level $k+2$ 
and three free fermions of signature $(-,+,+)$. 
\boldmath
\paragraph{$\slr$ from $\slc$} 
\unboldmath
To deal with the Lorentzian signature of the $\slr$ group manifold, 
it is convenient to decompose the $\slr$ representations according to its 
time-like elliptic subalgebra. It corresponds to the equivalence
\begin{equation}
\slr_k \sim  \frac{\slc|_k \times \text{U(1)}_{-k}}{\zi_k} \, ,
\label{cosdecomps}
\end{equation} 
where the right-hand side is written in terms of the 
super-coset $\slc$ and a time-like free boson $X^0$ at radius 
$\sqrt{2k}$. AdS$_3$ spacetime is the 
universal cover of the $\slr$ group manifold, obtained 
by taking a continuous $\zi$ orbifold instead of the 
discrete $\zi_k$. One can use this decomposition to define 
the Wick rotation of the AdS$_3$ \textsc{cft} in moduli space, 
see~\cite{Israel:2005ek} for details. 

To construct the closed string spectrum of the theory, we start with the 
partition function of $\slc \times $U(1) and implement the diagonal orbifold
action that defines the Euclidean AdS$_3$ \textsc{cft} 
in the standard way compatible with modular invariance. 
The conformal weights of the $\slc$ primaries in the \textsc{ns-ns} sector are given by
\begin{equation}
L_0 = -\frac{j(j-1)}{k-2} + \frac{(n+kw)^2}{4k} \quad , \qquad 
\bar{L}_0 = -\frac{j(j-1)}{k-2} + \frac{(n-kw)^2}{4k}\, . 
\label{cosetspec}
\end{equation}
$n$ and $w$ are respectively the momentum and winding around the cigar 
at infinity. The corresponding vertex operators are written as 
\begin{equation}
V^{j,\, \mathfrak{sl}_2/\mathfrak{u}_1}_{\frac{n-kw}{2} ; \frac{n+kw}{2}} (z,
\bar z)
\end{equation}
The vertex operators of the $\slr$ \textsc{wzw} are represented in the orbifold 
theory~(\ref{cosdecomps}) as follows:
\begin{equation}
V^{j,\, \mathfrak{sl}_2}_{m\, \bar m \, w_+ } = 
V^{j,\, \mathfrak{sl}_2/\mathfrak{u}_1}_{m-\frac{kw_+}{2} ; -\bar m + \frac{kw_+}{2}}
e^{i\sqrt{\frac{2}{k}} (m X^0 (z) + \bar m X^0 (\bar z) )}\, .
\end{equation}
with $m,\bar m \in \mathbb{R}$ (for the universal cover)
and $m-\bar m = n$. The sum $w^+$ of the winding numbers of the $\slc$ and
$U(1)$ theories is identified with the sector of {\it spectral flow}.

\paragraph{Representations and characters}
The characters of the $\slc$ super-coset
at level $k$ come in different categories corresponding to the classes of
irreducible representations of the $SL(2,\mathbb{R})$ algebra in the parent theory. In all cases
the quadratic Casimir of the representations is $c_2=-j(j-1)$. 

Firstly we consider \emph{continuous representations}, with $j = 1/2 + ip$,
$p \in \mathbb{R}^+$. The characters are denoted by
 $ch_c (p,m) \oao{a}{b}$, where the $\mathcal{N}=2$ superconformal
 $U(1)_R$ charge of the primary is $Q=2m/k$, $m \in \zi/2$. Explicitly they are given by:
\begin{equation}
ch_c (p,m;\tau,\nu) \oao{a}{b} = 
q^{\frac{p^2+m^2}{k}}e^{4i\pi\nu \frac{m}{k}} \frac{\vartheta \oao{a}{b} (\tau, \nu)}{\eta^3 (\tau)}\, .
\end{equation}

Then we have \emph{discrete representations} with $\nicefrac{1}{2} < j < \nicefrac{k+1}{2}$,
of characters $ch_d (j,r) \oao{a}{b}$, where the $\mathcal{N}=2$ $U(1)_R$
charge is $Q= (2j+2r+a)/k$, $r\in \zi$. The characters read: 
\begin{equation}
ch_d (j,r;\tau,\nu) \oao{a}{b} = q^{\frac{-(j-1/2)^2+(j+r+a/2)^2}{k}} 
e^{2i\pi\nu \frac{2j+2r+a}{k}} \frac{1}{1+(-)^b \, 
e^{2i\pi \nu} q^{1/2+r+a/2}} \frac{\vartheta \oao{a}{b} (\tau, \nu)}{\eta^3 (\tau)} .
\end{equation}

While the closed string spectrum in $\slr$ contains only discrete and
continuous representations, the spectrum of open strings attached to
localized D-branes is built on the {\it identity representation}.
The character for this identity representation
we denote by $ch_\mathbb{I} (r) \oao{a}{b}$. It is given by:
\begin{equation}
ch_\mathbb{I} (r;\tau,\nu) \oao{a}{b} =  \frac{(1-q)\
  q^{\frac{-1/4+(r+a/2)^2}{k}} 
e^{2i\pi\nu \frac{2r+a}{k}}}{\left( 1+(-)^b \, 
e^{2i\pi \nu} q^{1/2+r+a/2} \right)\left( 1+(-)^b \, e^{-2i\pi \nu} 
q^{1/2-r-a/2}\right)} \frac{\vartheta \oao{a}{b} (\tau, \nu)}{\eta^3 (\tau)}.
\end{equation}
These characters have the reflection symmetry
\begin{equation}
ch_\mathbb{I} \oao{-a}{-b} (-r;it) = ch_\mathbb{I} \oao{a}{b} (r;it)\, .
\label{reflsym} 
\end{equation}  
The primaries in the NS sector for this identity representation are as follows.
First we have the identity operator $|j=0,r=0\rangle \otimes
| 0\rangle_\textsc{ns}$. The other primary states are:
\begin{eqnarray*}
|r\rangle = \psi^{+}_{-\frac{1}{2}} |0\rangle_\textsc{ns} 
\otimes (J^{+}_{-1} )^{r-1} |0,0\rangle_\textsc{bos} \quad \text{for}\ r>0
\quad \text{with}\ 
&  L_0 & =  \frac{r^2}{k} +r - \frac{1}{2} \\
|r\rangle = \psi^{-}_{-\frac{1}{2}}|0\rangle_\textsc{ns} 
\otimes (J^{+}_{-1} )^{-r-1} |0,0\rangle_\textsc{bos} 
\quad \text{for}\ r<0
\quad \text{with}\ 
&L_0& = \frac{r^2}{k} -r - \frac{1}{2}\, .
\end{eqnarray*}

\section{From the annulus to closed strings emission}
\label{annulustransfo}
In this appendix we show how to get the mean number of emitted 
closed strings from the imaginary part of the annulus 
amplitude, given by eqn,~(\ref{imannul}). As quoted in the 
text the sum over $r$ is finite because the divergent 
terms from $\exp{-\pi t ( n-i\sqrt{\nicefrac{2}{k}}(r+b/2) )^2}$ 
are compensated with the weights in the $\slc$ characters. However 
it prevents from performing the modular transform to the closed 
string channel in a straightforward way. To do so we need to 
add a regulator to the amplitude~(\ref{imannul}): 
\begin{multline}
\text{Im} \, \mathcal{A}_{\epsilon}  =  \frac{1}{2} \int_0^\infty \frac{\di t}{2t}
\sum_{a \in \zi_2}(-)^b 
\chi^{0} (it )\ \frac{\vartheta \oao{b}{a}^3 (it )}{\eta^3 (it )} \sum_{\{ w^i\}} 
\frac{e^{-\pi t (R_i w^i)^2}}{\eta^3 (it )} 
 \sum_{r \in \zi} e^{-\pi \epsilon r^2}\ ch_\mathbb{I} \oao{b}{a} (r;it)
 \times  \\ \times \ 
\sum_{n=1}^{\infty}
 (-)^{a(n+1)}\, n \,  e^{-\pi t \left( n-i\sqrt{\frac{2}{k}}(r+b/2) \right)^2}  .
\label{imannulreg}
\end{multline} 
Then we can modular transform the various factors and get first 
a contribution from the continuous representations, with $s=1/t$:
\begin{multline}
\frac{\pi}{k^2 \prod_i R_i} \int_0^\infty \di s \sum_{a \in \zi_2}(-)^b 
\sum_{\tilde \jmath} \sin \frac{\pi (2\tilde \jmath +1)}{k}  \chi^{\tilde \jmath} (is )
\frac{\vartheta \oao{a}{b}^3 (is )}{\eta^3 (is )}
\sum_{\{ n_i\}} 
\frac{e^{-\pi s (n_i/R_i)^2}}{\eta^3 (is )} \ \times  \\ \times \  
\int \di P \di \mu \frac{\sinh 2\pi P \sinh 2\pi P/k}{\cosh 2\pi P + \cos 2\pi
  \mu} ch_c (P,\mu;is) \sum_{r\in \zi} e^{-\pi \epsilon r^2}
e^{-\frac{4i\pi \mu (r+b/2)}{k}} \ \times  \\ \times \ 
\sum_{n=1}^{\infty}
 (-)^{a(n+1)}\,  \int \di p_0 \ e^{-\frac{\pi s p_0^2}{2k}} 
e^{2i\pi p_0 (\frac{n}{\sqrt{2k}}-\frac{i(r+b/2)}{k})} \, .
\end{multline}
Now we expand the contribution of the various characters 
in the partition function. We have schematically, e.g. in the \textsc{ns}  
sector:
\begin{multline}
 \chi^{\tilde \jmath} (is )  
\frac{\vartheta \oao{0}{b}^3 (is )}{\eta^3 (is )} 
\frac{e^{-\pi s  
\sum_i \left(\nicefrac{n_i}{R_i}\right)^2}}{\eta^3 (is )}\ 
ch_{c} \oao{0}{b} (P,\mu;is)  \\ 
\sim \sum_{N ; \, \textsc{ns}}  
(-)^{bF} D(N) e^{-2\pi s \left( \frac{\tilde \jmath(\tilde \jmath+1)}{k} +
    \frac{P^2+\nicefrac{1}{4}}{k}+ \frac{\mu^2}{k} + \frac{(n_i/R_i)^2}{2}+ N 
- \frac{1}{2}\right)} \, .
\end{multline}
where $D(N)$ is the density of states at 
oscillator number $N$,  and $F$ the worldsheet fermion number for a given string
state. Of course one would need to expand more rigorously the various
characters and to keep track of the null vectors of $SU(2)_k$. After
integrating over the Schwinger parameter $s$ we get
\begin{multline}
\frac{1}{2k^2 \prod_i R_i} \sum_{a,b} \sum_{N} \sum_{\tilde \jmath}  \sin \frac{\pi (2\tilde
  \jmath +1)}{k}  \sum_{n=1}^{\infty} n 
(-)^{a(n+1)+bF} D(N)\ \times\\ \times \  \int \di \mu \di P 
\frac{\sinh 2\pi P \sinh 2\pi P/k}{\cosh 2\pi P + \cos 2\pi
  \mu} e^{-\frac{2i\pi \mu b}{k}}\ \times \\  
 \int \di p_0  \frac{e^{2i\pi p_0 (\frac{n 
  }{\sqrt{2k}}+\frac{ib}{2k})} 
}{\frac{p_0^2}{4k} + \frac{\mu^2}{k} +  \frac{\tilde \jmath(\tilde \jmath+1)}{k} +
    \frac{P^2+\nicefrac{1}{4}}{k}+ \frac{\mu^2}{k} + \frac{(n_i/R_i)^2}{2}+ N 
 - \frac{a-1}{2}} \sum_{r \in \zi}  e^{-\pi \epsilon r^2+2\pi p_0 \frac{r}{k}-\frac{4i\pi \mu r}{k}}
\end{multline}
The integral over $p_0$ closes in the upper-half plane and gets a contribution 
from the simple pole at $p_0 = 2i\sqrt{\mu^2+k\Delta}$ with 
\begin{equation}
\Delta = \frac{\tilde \jmath(\tilde \jmath+1)}{k} +
    \frac{P^2+\nicefrac{1}{4}}{k}+ \frac{\mu^2}{k} + \frac{(n_i/R_i)^2}{2}+ N 
+ \frac{a-1}{2} \, .
\end{equation}
We get 
\begin{multline}
\frac{\pi}{k \prod_i R_i} \sum_{a,b} \sum_{N} \sum_{\tilde \jmath}  \sin \frac{\pi (2\tilde
  \jmath +1)}{k}  \sum_{n=1}^{\infty} n 
(-)^{a(n+1)+bF} D(N) \ \times \\ \times \  \int \di \mu \di P 
\frac{\sinh 2\pi P \sinh 2\pi P/k}{\cosh 2\pi P + \cos 2\pi
  \mu}   e^{-\frac{2i\pi \mu b}{k}} \ \times \\ \times \ 
\int \di p_0 \ \frac{e^{-4\pi \sqrt{\mu^2+k\Delta} (\frac{n 
  }{\sqrt{2k}}+\frac{ib}{2k})} 
}{\sqrt{\mu^2+k\Delta}} \sum_{r \in \zi}  e^{-\pi \epsilon r^2}
e^{\frac{4i\pi r}{k}(\sqrt{\mu^2+k\Delta} -\mu)}\, .
\end{multline}
Now we can take the limit $\epsilon \to 0$ and rewrite this as 
\begin{multline}
\frac{\pi}{2\prod_i R_i} \sum_{a,b} \sum_{N} \sum_{\tilde \jmath}  \sin \frac{\pi (2\tilde
  \jmath +1)}{k}  \sum_{n=1}^{\infty} n 
(-)^{a(n+1)+bF} D(N) \ \times \\ \times \   \int \di \mu \di P 
\frac{\sinh 2\pi P \sinh 2\pi P/k}{\cosh 2\pi P + \cos 2\pi
  \mu} e^{-2i\pi b (\mu -  \sqrt{\mu^2+k\Delta})}
 \ \frac{e^{-4\pi \sqrt{\mu^2+k\Delta} \frac{n 
  }{\sqrt{2k}}} 
}{\sqrt{\mu^2+k\Delta}}  \ \times \\ \times \
\sum_{w \in \zi} \delta (\mu - \sqrt{\mu^2+k\Delta}+\frac{kw}{2}) \, .
\end{multline}
And finally integrating over $\mu$ gives
\begin{multline}
\frac{\pi}{k\prod_i R_i} \sum_{a,b} \sum_{N} \sum_{\tilde \jmath}  \sin \frac{\pi (2\tilde
  \jmath +1)}{k}  
(-)^{a+bF} D(N) \ \times \\ \times \   \sum_{w \neq 0} \frac{1}{|w|}  \int \di P 
\frac{\sinh 2\pi P \sinh 2\pi P/k}{\cosh 2\pi P + \cos \pi
  (\frac{2\Delta}{w}-\frac{kw}{2})}   \sum_{n=1}^{\infty} 
(-)^{an}
n \ e^{-\frac{2\pi n}{\sqrt{2k}} \left( -\frac{kw}{2}+\frac{2\Delta}{w}\right)} \\
= \frac{1}{4k\prod_i R_i} \sum_{a,b} \sum_{N} \sum_{\tilde \jmath}  \sin \frac{\pi (2\tilde
  \jmath +1)}{k}  
(-)^{bF}\ \times \\ \times \ D(N)  \sum_{w \neq 0} \frac{1}{|w|} \int \di P  
\frac{\sinh 2\pi P \sinh 2\pi P/k}{\cosh 2\pi P + \cos \pi
  (\frac{kw}{2}-\frac{2\Delta}{w})}  
 \frac{(-)^a}{\sinh^2 \pi \left( \frac{
  \frac{kw}{2}+\frac{2\Delta}{w}}{\sqrt{2k}} + \frac{ia}{2}\right)}\, ,
\end{multline}
which is the same as~(\ref{prodlongexpl}). The discrete representations
contribution is obtained similarly.

\section{Long strings emission for bosonic AdS$_3$ backgrounds} 
\label{bosprod}
In this appendix we discuss the power-law corrections to t
long string emission~(\ref{prodlongexpl}) in 
non-critical bosonic string backgrounds of the form 
$\slr |_{k} \times \mathbb{R}^d$. In this case the counting 
of states can be done quite explicitly. Cancellation of the conformal anomaly in this 
background requires 
\begin{equation}
k=2+\frac{6}{23-d}.
\label{cchargebos}
\end{equation}
Taking into account the differences between the superstring and bosonic string cases, 
the imaginary part of the annulus amplitude for long strings reads 
\begin{multline}
\bar{\mathcal{N}}_c \sim \int \di^d p \ 
 \int \di P   \sum_{N}  D(N) 
\sum_{w \neq 0} \frac{1}{|w|}\ 
\frac{ \sinh 2\pi P  \sinh \frac{2\pi P}{k-2}}{\cosh 2\pi P + 
\cos \pi (kw -E)}   
 \frac{1}{\sinh^2 \frac{\pi E}{\sqrt{k}}}\, ,
\end{multline}
with 
\begin{equation}
E = \frac{kw}{2} + \frac{2}{w} \left[ \frac{P^2+1/4}{k-2} + \frac{\mathbf{p}^2}{2}+N-1 \right]
\end{equation}
At high energies, this amplitude has the asymptotics
\begin{equation}
\bar{\mathcal{N}}_c \sim 
\sum_{w \neq 0} \frac{1}{|w|} e^{-\pi \sqrt{k} w}
\left[\int \di P \ e^{\frac{2\pi P}{k-2}-\frac{4\pi}{w\sqrt{k}} \frac{P^2}{k-2}}\right]
\left[\int \di p\ e^{-\frac{2\pi}{w\sqrt{k}} p^2}\right]^d
\sum_N D(N) e^{-\frac{4\pi}{w} \sqrt{\frac{2}{k}}N}\, .
\end{equation}
Let's first integrate over the $\mathbb{R}^d$ momentum (bringing a factor $w^{d/2}$) and over 
the AdS$_3$ radial momentum $P$:
\begin{equation}
\bar{\mathcal{N}}_c \sim 
\sum_{w=1}^\infty w^{\frac{d-1}{2}} e^{-2\pi \sqrt{k}\left( 1-\frac{1}{4(k-2)} \right) w}
\sum_N D(N) e^{-\frac{4\pi N}{w\sqrt{k}} }
\end{equation}
The asymptotic density of states $D(N)$ corresponds to the counting of oscillator modes 
for $d+1$ bosons, in the light-cone gauge.
\begin{equation}
\frac{q^{(d+1)/24}}{\eta^{d+1} (\tau)} = \left(\prod\limits_{n=1}^{\infty} (1-q^n) \right)^{-(d+1)}= 
\sum_N D(N) q^N \, . 
\end{equation}
One can use the generalized Ramanujan and Hardy formula:
\begin{equation}
D(N) \stackrel{N \gg 1}{\sim} \frac{1}{\sqrt{2}} \left(\frac{d+1}{24}\right)^{\frac{d+2}{4}} N^{-\frac{d}{4}-1} 
e^{2\pi \sqrt{\frac{(d+1)N}{6}}}\, . 
\end{equation}
Then one replaces the sum over $N$ by an integral:
\begin{multline}
\sum_N N^{-\frac{d+4}{4}}  e^{2\pi \sqrt{\frac{(d+1)N}{6}}} e^{-\frac{4\pi N}{w\sqrt{k}} } 
\sim \int \frac{\di N}{N} N^{-d/4} e^{2\pi \sqrt{\frac{(d+1)N}{6}}-\frac{4\pi N}{w\sqrt{k}} } \\
\sim w^{-d/4} \int \frac{\di x}{x^{1+d/2}} e^{-x^2+\sqrt{\frac{(d+1)\sqrt{k}\pi w}{6}}\, x} 
\sim w^{-\frac{d+1}{2}} e^{\pi \sqrt{k} \frac{d+1}{24} w}
\end{multline}
Therefore using~(\ref{cchargebos}) all the exponential terms cancel. We remain with the power-law corrections, that 
don't depend on $d$:
\begin{equation}
\bar{\mathcal{N}}_c \sim \sum_{w=1}^{\infty} \frac{1}{w}
\end{equation}
which is log-divergent. This demonstrates that all the non-critical bosonic AdS$_3$ backgrounds have a common 
divergence in long strings emission. 

\bibliography{biblitachyon}

\end{document}